\documentclass[twocolumn,superscriptaddress,floatfix]{revtex4-1}

\usepackage{graphicx,amsmath,units,hyperref,upgreek}

\begin{document}



\title{Revealing the dynamics of ultrarelativistic non-equilibrium many-electron systems with phase space tomography}



\author{Stefan Funkner}
\email{stefan.funkner@kit.edu}
\affiliation{Karlsruhe Institute of Technology, 76344 Eggenstein-Leopoldshafen, Germany}
\author{Gudrun Niehues}
\affiliation{Karlsruhe Institute of Technology, 76344 Eggenstein-Leopoldshafen, Germany}
\author{Michael J. Nasse}
\affiliation{Karlsruhe Institute of Technology, 76344 Eggenstein-Leopoldshafen, Germany}
\author{Erik Br\"{u}ndermann}
\affiliation{Karlsruhe Institute of Technology, 76344 Eggenstein-Leopoldshafen, Germany}
\author{Michele Caselle}
\affiliation{Karlsruhe Institute of Technology, 76344 Eggenstein-Leopoldshafen, Germany}
\author{Benjamin Kehrer}
\affiliation{Karlsruhe Institute of Technology, 76344 Eggenstein-Leopoldshafen, Germany}
\author{Lorenzo Rota}
\altaffiliation[Present address:]{ SLAC, Stanford, CA, USA}
\affiliation{Karlsruhe Institute of Technology, 76344 Eggenstein-Leopoldshafen, Germany}
\author{Patrik Sch\"{o}nfeldt}
\altaffiliation[Present address:]{ DLR-VE, Oldenburg, Germany}
\affiliation{Karlsruhe Institute of Technology, 76344 Eggenstein-Leopoldshafen, Germany}
\author{Marcel Schuh}
\affiliation{Karlsruhe Institute of Technology, 76344 Eggenstein-Leopoldshafen, Germany}
\author{Bernd Steffen}
\affiliation{Deutsches Elektronen Synchrotron, 22607 Hamburg, Germany}
\author{Johannes L. Steinmann}
\altaffiliation[Present address:]{ APS, Lemont, USA}
\affiliation{Karlsruhe Institute of Technology, 76344 Eggenstein-Leopoldshafen, Germany}
\author{Marc Weber}
\affiliation{Karlsruhe Institute of Technology, 76344 Eggenstein-Leopoldshafen, Germany}
\author{Anke-Susanne M\"{u}ller}
\affiliation{Karlsruhe Institute of Technology, 76344 Eggenstein-Leopoldshafen, Germany}


\date{\today}

\begin{abstract}
The description of physical processes with many-particle systems is a key approach to the modeling of countless physical systems. In storage rings, where ultrarelativistic particles are agglomerated in dense bunches, the measurement of their phase-space distribution (PSD) is of paramount importance: at any time the PSD not only determines the complete space-time evolution but also provides fundamental performance characteristics for storage ring operation. Here, we demonstrate a non-destructive tomographic imaging technique for the 2D longitudinal PSD of ultrarelativistic electron bunches. For this purpose, we utilize a unique setup, which streams turn-by-turn near-field measurements of bunch profiles at MHz repetition rates. To demonstrate the feasibility of our method, we induce a non-equilibrium state and show, that the PSD microstructuring as well as the PSD dynamics can be observed in great detail with an unprecedented resolution. Our approach offers a pathway to control ultrashort bunches and supports, as one example, the development of compact accelerators with low energy footprints.
\end{abstract}

\maketitle

\section{Introduction} 

The modeling of systems composed of a large number of interacting particles are of utmost importance in physics \cite{Arnold2004} with applications reaching from molecular dynamics simulations \cite{Alder1959}, the description of traffic dynamics \cite{Helbing2001} up to the treatment of quantized many-particle systems \cite{Negele2018}. \\
One of the idealized model systems is the many-electron system. In applications, many-electron systems can be affected by relativistic effects, quantized emission of photons and coherence, if they are spatially compressed. For example, relativistic free electrons are utilized in particle accelerators for high-energy physics experiments and the generation of synchrotron radiation. In the latter case, the fact that accelerated ultrarelativistic electrons produce broadband emission from the microwave to the hard x-ray range is harnessed in evermore advanced synchrotron light sources throughout the word for widespread applications in science, industry and medicine. \\
For such applications, accelerator-based light sources need to provide stable emission. This is achieved by storage rings, where the particles circulate in pulsed structures consisting of electron bunches. If sufficiently compressed, these bunches emit coherent synchrotron radiation (CSR) at wavelengths corresponding to the Fourier transform of their longitudinal density profile \cite{Schwinger1996,Schiff1946,PhysRevAccelBeams.19.020704}.  Compared to incoherent radiation, CSR can be several orders of magnitude higher in intensity, which makes it particularly interesting for user applications. Hence, to provide an ever-increasing brilliance at frequencies in the THz range and above \cite{Tammaro2015}, scientists aim to increase the electron density by compressing the bunch as short as possible. \\
However, there is a trade-off to this optimization: above a certain density threshold the collective radiation field exerts significant forces on the particles leading to a buildup and self-amplification of substructures on the bunch profile. This formation of substructures of an electron bunch with the associated dynamics is called microbunching instability (MBI). As a consequence, irregular bursts of CSR are emitted, which can at short wavelengths be even more intense than the CSR during stable bunch formation at an electron density below the stability threshold \cite{Stupakov2002}. The MBI is the main limiting factor for the generation of stable and intense CSR in storage rings run in short bunch mode. The MBI is furthermore considered as a natural process occurring during the emission of solar flares \cite{Kaufmann2006}.\\ 
During the MBI the electron bunch might be re\-gard\-ed as a non-equilibrium thermodynamic sys\-tem with a steady flow of energy exhibiting rich structural and dynamic self-organized patterns similar to other non-equi\-li\-brium thermodynamic systems, such as the Rayleigh-B{\'e}rnard convection, Turing instability or the Belousov-Zhabotinsky reaction \cite{Cross1993}. Here, a deep understanding of the physical laws determining the development of the electron bunch density bears the potential to further control or stabilize the occurrence of the CSR bursts, which then could provide a bright THz radiation source for user applications \cite{Evain2019}. The details of the dynamics during the MBI are, however, still unclear and subject of intense research efforts  \cite{Roussel2015a, wustefeld2010coherent, Abo-Bakr2003, Shields2012, warnock2000general, Bane2010, Evain2012, Schonfeldt2017, Brosi2019, CARR2001387, Byrd2002ObservationOB, Karantzoulis2010, Steinmann2018}.\\   
With respect to accelerator diagnostics, measuring the phase space distribution (PSD) of the electron bunch is key to a complete characterization of its physical state. While on the theoretical side, phase space dynamics can be simulated with great detail (for example by numerically solving the corresponding Vlasov-Fokker-Planck equation \cite{Schonfeldt2017,warnock2000general}), for non-destructive experiments with electron bunches the PSD remains an elusive quantity. Early ideas by Hancock et.~al.~\cite{Hancock2000} for the reconstruction of the longitudinal PSD of proton bunches and other concepts for heavy ion colliders \cite{Montag2002}, linear accelerators \cite{Yakimenko2003, Hock2013} as well as theoretical considerations \cite{Michelotti2003}, did not mature into a diagnostic tool for electron storage rings, where challenges from the high revolution frequencies of rings and fading out of information about substructure in the far-field have to be overcome. Hence, many interpretations of experimental results can only be taken as indirect conclusions about the state of the PSD of the electron bunches.\\
In this paper, we describe how this diagnostics gap can be closed using single-shot electro-optical (EO) sampling of the electron bunch near-field in combination with turn-by-turn measurements at MHz repetition rates using an ultra-fast line-array detection system \cite{Rota2018}. The longitudinal PSD not only determines the emission spectrum of the bunch, but also provides insights in other fundamental properties such as bunch length, energy spread or, more sophisticated, relativistic intra-bunch interactions. Throughout the paper we restrict our attention to the longitudinal PSD. We therefore consider the spatial coordinate in the direction of the bunch motion and approximate the bunch as one-dimensional line charge density. 
\section{Reconstruction method}
\begin{figure}
\includegraphics[width=\linewidth]{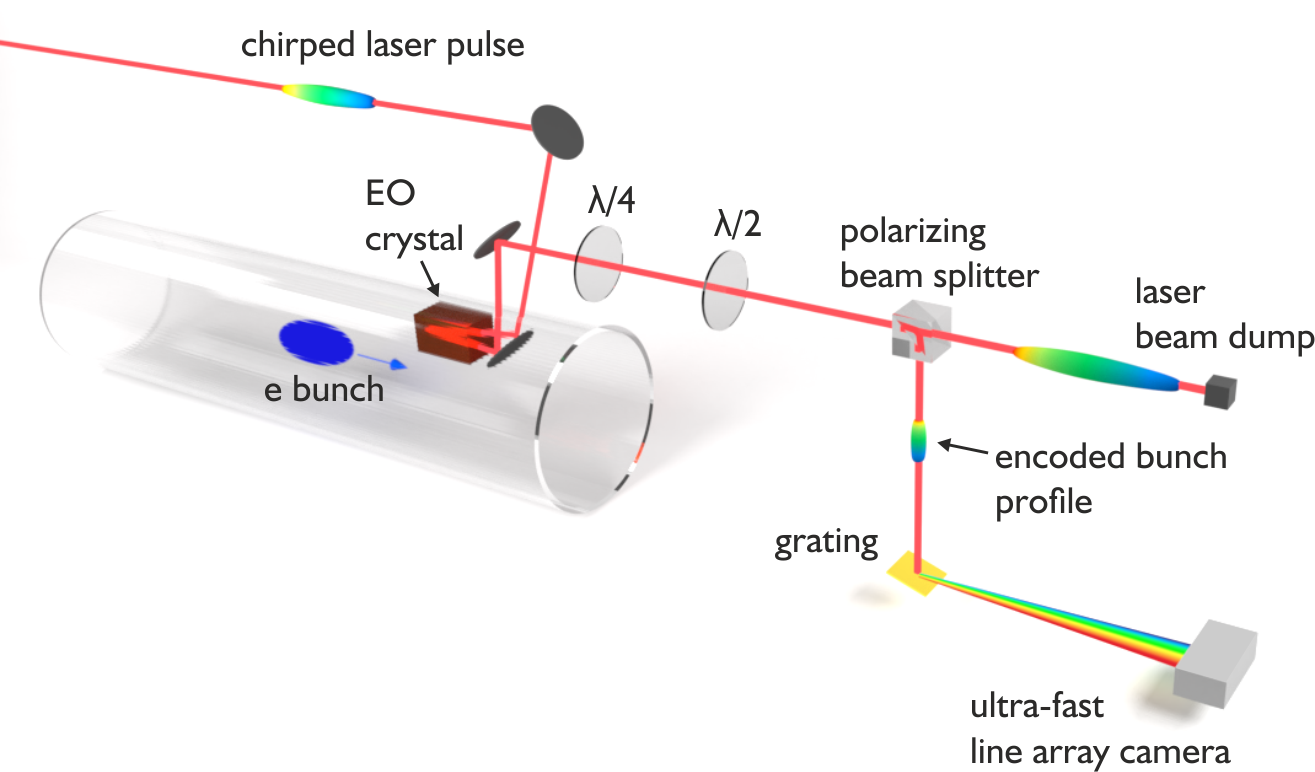}
\caption{\label{setup} The setup to measure charge density profiles of electron bunches in a storage ring with single-shot electro-optical sampling of the near-field is shown (see \cite{Casalbuoni2008, Muller2012, Funkner2019}). The closed metallic beam pipe of the accelerator is here indicated as a transparent cylinder.}
\end{figure}
\begin{figure*}
\includegraphics[width=\linewidth]{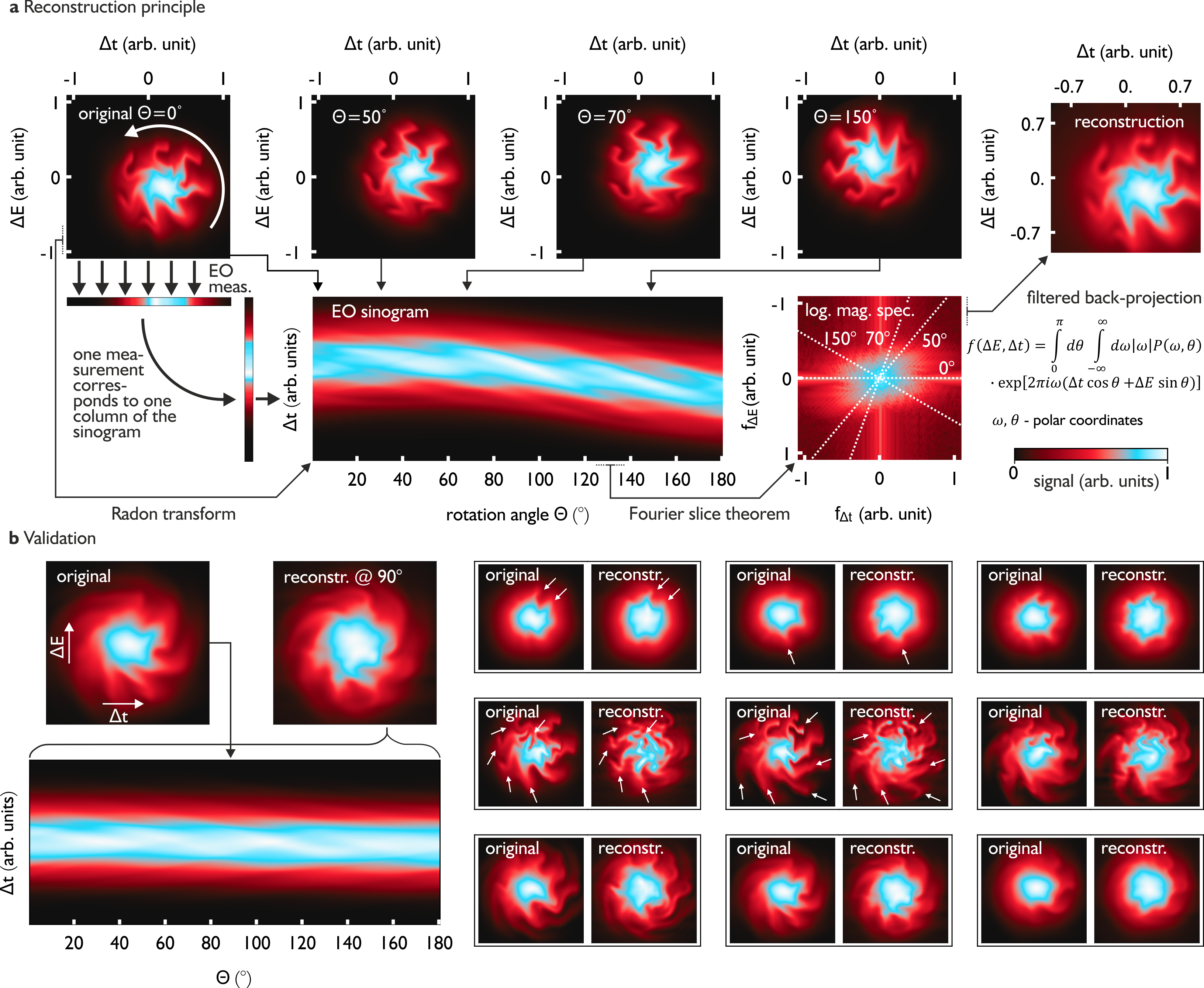}
\caption{\label{reconstruction} a) Reconstruction principle of the PSD distribution with electro-optical sampling. Top: we assume a rigid rotation. Shown are the PSD configurations at different rotation angles and the final reconstruction. Bottom: the EO measurement (projection on the time axis) of the different PSD configurations can be interpreted as a Radon transform. Each PSD configuration is represented by a single column in the corresponding sinogram. The Radon transform is connected to the 2D Fourier transform of the original image via the Fourier slicing theorem. The original PSD can be reconstructed with a filtered back-projection.  b) Validation of the reconstruction algorithm with simulated data obtained from the Vlasov-Fokker-Planck solver Inovesa \cite{Schonfeldt2017}. We used a standard ramp filter to avoid the blurring effect of the back-projection \cite{Buzug2008} in combination with a high frequency cut-off at a relative frequency of 0.1 to reduce the exaggeration of edges. Left side: comparison between the original and reconstructed PSD. The ``original'' PSD is evaluated at the $90^\circ$ frame, while for the reconstruction the complete sinogram has to be taken into the account. Right side: several comparisons during different times of a bursting cycle. Even in situations with prominent substructures, where bunch self-interaction is expected to be strong, the reconstruction shows a very good agreement with the original PSD, even in fine details.}
\end{figure*}
\begin{figure*}
\includegraphics[width=\linewidth]{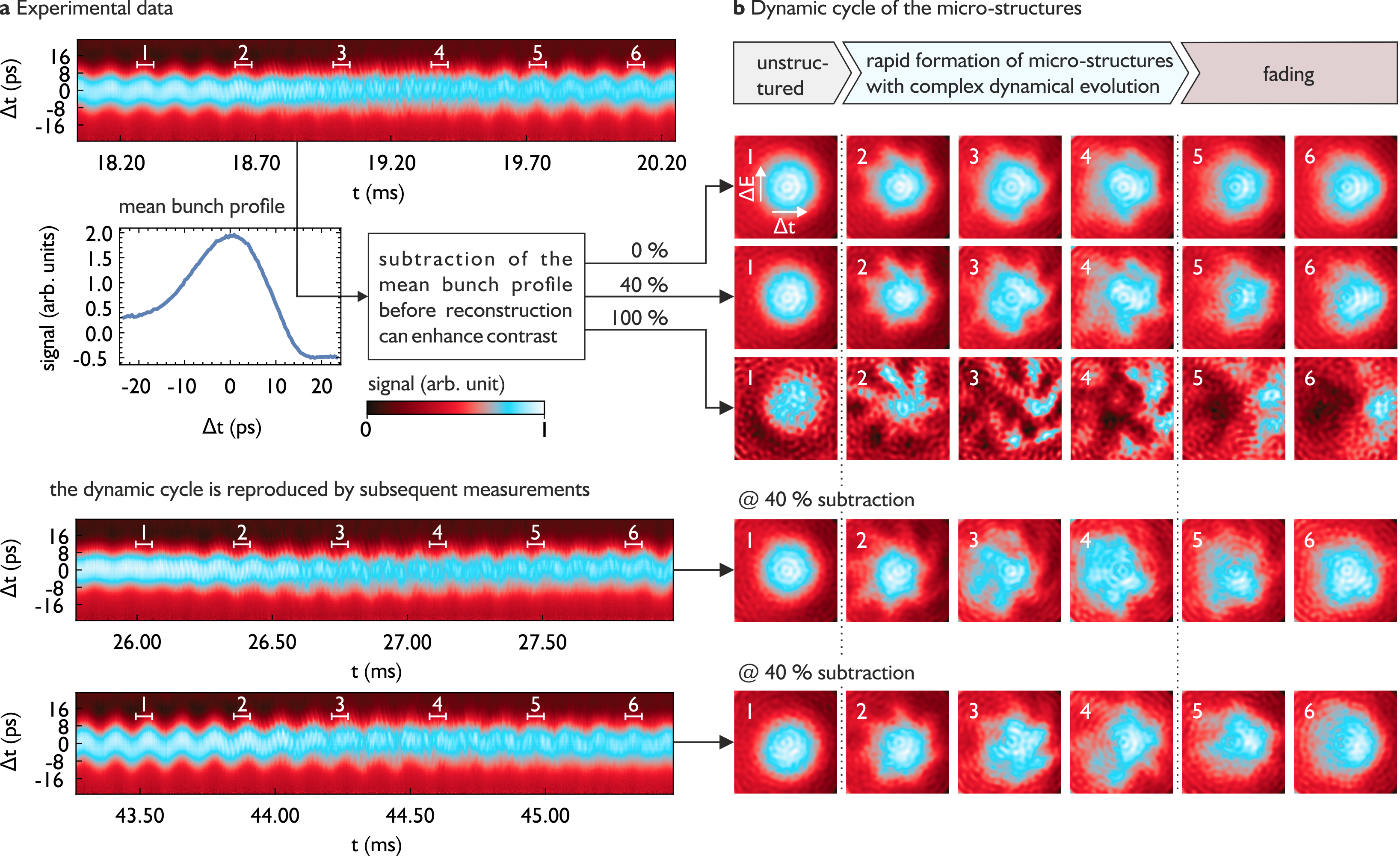}
\caption{\label{experiment} a) Experimental data: three revolution plots of 6000 consecutive measurements (corresponding to 18 synchrotron oscillations) of the electron bunch charge density taken from a longer data set with 500,000 measurements ($\sim$ 0.18 s). For each revolution plot, the measurements are concatenated from left to right. The value for $\Delta$t is determined by the pixel number of the line array, calibration measurements \cite{Funkner2019} and the mean bunch profile position. b) Corresponding reconstructions of the PSD for different time intervals. The numbers refer to the time intervals marked in the revolution plots on the left side, respectively. Upper side: reconstructions for different subtraction percentages of the mean bunch profile are compared for the first revolution plot. The dynamic cycle of the microbunching instability is repeated over and over again, as it can be seen by comparing the rows of the reconstructed PSDs for the different revolution plots. To provide the optimal contrast, we normalized the data for every image independently.}
\end{figure*}
The working principle of our setup is visualized in Fig.~\ref{setup} (see \cite{Funkner2019} for a detailed description): to measure the charge density of a relativistic bunch we overlap broadband linearly polarized chirped laser pulses from a fiber laser with the Coulomb field of the electron bunch in a gallium phosphide EO crystal. The timing is adjusted by a phase-locked loop synchronizing the laser repetition rate to the revolution frequency of the storage ring. Due to the relativistic motion of the bunch, the Coulomb field of an individual relativistic electron is highly compressed perpendicular to its motion in a radial pancake-like structure. Consequently, the time-dependent Coulomb field of the electron bunch leaked into the EO crystal is proportional to the charge density profile. Due to the chirp in the linearly polarized beam, the charge density profile is imprinted on the laser pulse as a wavelength dependent change of the polarization. \\
In the next stage, this information is transformed into a wavelength dependent intensity profile by a polarizing cube beam splitter. In this process, the wavelength encodes the time-dependence. The profile is finally decoded by a grating and read out by an ultra-fast line array camera \cite{Rota2018} in the spectral domain. By comparing the intensity profile of the laser pulse to a reference measurement, taken without the presence of electrons, we can directly deduce the longitudinal charge profile of the electron bunch \cite{Funkner2019}. \\
From the bunch profile measurements, we can reconstruct the PSD using the following approximation: during the time of half a synchrotron oscillation period (in our case $\sim$ \unit[61]{$\upmu$s}), we assume that the change of the microstructures is small. Then the resulting dynamics can be described by a well-known rotation of the PSD \cite{Evain2012}, which is a result of the phase focusing of the electron bunch to the radio frequency phase of the storage ring \cite{Wille2001}. In other words, we approximate the dynamics of the PSD by a rigid rotation with a rotation period of $\sim$ \unit[122]{{$\upmu$}s}. \\
In Fig.\,\ref{reconstruction}\,a, we show a typical configuration of the PSD taken from a simulation of the electron bunch dy\-na\-mics based on the Vlasov-Fokker-Planck equation using Ino\-ve\-sa, a simulation program developed at the Karlsruhe Institute of Technology (KIT) \cite{Schonfeldt2017}. With respect to the phase space, an EO near-field measurement can be interpreted as a projection of PSD onto the axis related to a generalized coordinate (in our case we measure the arrival time). During consecutive measurements, performed on a turn-by-turn basis by our setup, the PSD of the electron bunch rotates, so that the experiment can be interpreted as a tomographic measurement of the electron bunch PSD. If we concatenate consecutive measurements taken during half a rotation period we obtain, in this interpretation, the Radon transform of the PSD resulting in the so-called sinogram (see Fig.~\ref{reconstruction}\,a bottom). The sinogram can be related to the 2D Fourier transform of the original image (or PSD) via the Fourier slice theorem \cite{Jahne2005}. The reconstruction of the PSD is then performed with a filtered back-projection \cite{Jahne2005}, a well-known procedure widely used for example in medicine for computed tomography scans.  \\
The approximation of the PSD dynamics as a rigid rotation is an idealization made to motivate the reconstruction algorithm. This situation might be different especially above the stability threshold, where the substructures actually form because of deviations from the rigid rotation model \cite{PhysRevAccelBeams.19.020704}. Hence, to validate our approach for the condition present during the experiments we use Inovesa simulations. The simulations provide a PSD during every time step, so that we can calculate the EO sinogram (which is a simple projection) and compare the corresponding reconstructed PSD to the original simulated PSD. In Fig.~\ref{reconstruction}\,b, we show that even though the PSD dynamics can be complicated during the MBI due to collective effects of the microstructures, the filtered back-projection can not only reconstruct the PSD on a coarse grained level, but also resolve subtleties of the microstructures (we marked a few of them with white arrows).\\
To demonstrate this principle experimentally, we measured profiles of a single electron bunch during the MBI at the Karlsruhe Research Accelerator (KARA). During the experiments, the bunch contained roughly  $2\cdot 10^9$ electrons and was accelerated to a highly relativistic motion with a Lorentz factor $\gamma \approx 2500$. To induce the MBI, we compressed the longitudinal bunch size with the low-$\alpha_c$ mode of the storage ring to a few picoseconds. We then observed the development of the bunch profile at a repetition rate of \unit[2.72]{MHz} corresponding to the revolution time of a bunch in the storage ring. \\
In Fig.~\ref{experiment}a, we show three different sections each consisting of 6000 successive profile measurements stacked from left to right. The sections are part of a larger data set with 500,000 turn-by-turn measurements. The representation of the sections in Fig.~\ref{experiment}a is also referred to as ``revolution plot'' (see \cite{Funkner2019}). In each displayed revolution plot, the noise-induced \cite{Ormond} coherent synchrotron oscillation appears as a clearly visible center of mass oscillation of the bunch profile. \\
We utilize this effect to determine the synchrotron oscillation period, which is need\-ed to extract the sinograms for PSD reconstruction: if we consider a horizontal cross section in the revolution plots, the charge density fluctuates as a consequence of the synchrotron oscillation. Hence, the frequency of the synchrotron oscillation can be conveniently determined from the data set using a Fourier transform. To provide a precise estimation we used all bunch profiles of the data set and averaged the resulting spectrum along the horizontal cross sections.  In the case shown here, the synchrotron frequency is estimated to be \unit[8.28]{kHz} corresponding to 328 turns. Therefore, roughly 18 synchrotron oscillations are displayed for each revolution plot in Fig.\,\ref{experiment}\,a. \\
From the determination of the syn\-chro\-tron oscillation frequency, we conclude that a PSD reconstruction can be performed from a sinogram consisting of about 328/2=164 consecutive measurements (the angle resolution is about 1.1$^\circ $), i.e.\ a sinogram corresponds to a revolution plot with exact 164 measurements. In Fig.\,\ref{experiment}\,b, we show reconstructions of the PSD during different times. For the calculations we used a filtered back-projection with a ramp filter \cite{Buzug2008}. Compared to the simulations, we have a bunch profile sampling rate that is approximately two times lower, so that we doubled the high frequency cut-off for the filtered back-projection from 0.1 to 0.2 of the highest frequency.  The time intervals used for every reconstruction are indicated in each revolution plot in Fig.\,\ref{experiment}\,a as white horizontal lines. Each of these short intervals on its own can be regarded as a sinogram of the PSD. \\
In general, the observed microstructuring of the bunch profile is small compared to the overall profile. For the top revolution plot in Fig.\,\ref{experiment}, we show how the contrast of the PSD reconstructions can be enhanced: we calculate the mean bunch profile for each revolution plot and subtract a scaled version from every bunch profile before applying the filtered back-projection. Such a procedure might introduce artifacts and should be handled carefully. A comparison for different scaling factors is visible at top Fig.\,\ref{experiment}\,b. We note that a 40~\% subtraction can improve the contrast significantly without introducing additional features (deduced from a careful comparison with the PSD at 0~\% subtraction) as it might be the case when subtracting 100~\% of the mean bunch profile. At this point, we emphasize that the measured PSDs in Fig.\,\ref{experiment}\,a are remarkably similar to the simulated ones in Fig.\,\ref{reconstruction}\,b.\\
\begin{figure}
\includegraphics[width=\linewidth]{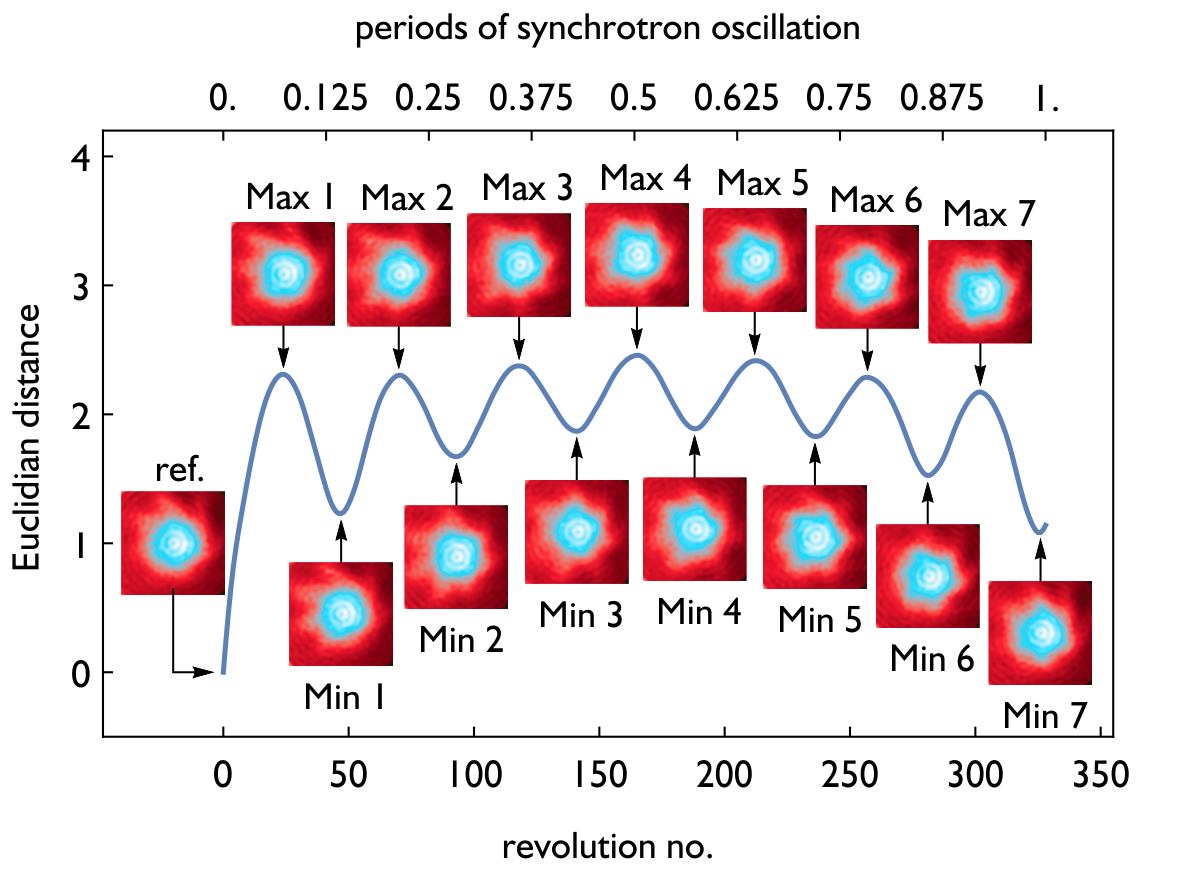}
\caption{\label{eudistance} The Euclidian distance between a reference PSD and each PSD from the subsequent data is shown (in blue). The PSD configurations at the minima and maxima as well as the reference PSD are displayed as insets. The Euclidian distance follows an oscillatory pattern as expected from a complete rigid rotation of a PSD structure with 7-fold symmetry.}
\end{figure}
In the next step, we use the measured PSDs, to analyze the development and the structure of the electron bunch distribution in phase-space. We first analyze the PSD reconstruction from the revolution plot in Fig.\,\ref{experiment}\,a at the top: at the beginning of the measurement, substructures due to the MBI are not visible in the revolution plot at time interval 1. Hence, the PSD, shown in the first column of Fig.\,\ref{experiment}\,b, corresponds approximately to a 2D Gaussian-like distribution. In the further development, diagonal stripes appear in the revolution plot. Here, the reconstruction of the PSD in Fig.\,\ref{experiment}\,b at column 2 shows a star-like structuring with a symmetry center roughly located at the center of mass position. Afterwards, the structure of the PSD during time intervals 3 and 4 undergoes a complex dynamic development and the stripe pattern becomes irregular. Finally, during the time intervals 5 and 6, the stripe pattern becomes less visible and the PSD images correspond to Gaussian distributions overlaid with weak substructures. This relaxation during which the substructures disappear takes rather long in comparison to the fast appearance of the substructures in the beginning. The overall process shown here retraces a typical excitation and relaxation pattern during the MBI (see \cite{Kehrer2018}). This  is underlined by the reproduced dynamics demonstrated by the reconstructions for the two other revolution plots in Fig.\,\ref{experiment}\,a. \\
While this shows how the overall dynamics can be easily analyzed with the reconstructed PSDs, it is also possible to extract information about the dynamical development of structural features of the PSD: as an example, we calculated all PSDs during one complete synchrotron oscillation period starting at \unit[18.74]{ms} of the measured data set (around the second step of the dynamic cycle in the first revolution plot in Fig.\,\ref{experiment}), where a star-like shape with a 7-fold symmetry is observed. \\ 
In the next step, we investigate the correlation of every image with the reference frame obtained from the first 164 revolutions using a simple Euclidean distance \cite{Nakhmani2013, LiweiWang2005}. The result, shown in Fig.\,\ref{eudistance}, features a well-defined oscillatory pattern. For the interpretation of the results it is important to note, that the reconstructed PSDs from time intervals up to half a synchrotron period after the reference frame are statistically dependent. This is due to the overlapping data used for the reconstructions (i.e. a sliding window is used for the turn-by-turn reconstructions). Therefore, the correlation should be investigated for more than half a synchrotron oscillation period, such as in Fig.\,\ref{eudistance}, where the oscillatory pattern is recorded for a complete synchrotron oscillation period. \\
While a distinct reconstruction provides structural in\-for\-ma\-tion, the correlation of the PSD over time adds dynamic information, which is different from a self-correlation of a single PSD measurement. Here, the clear occurrence of seven minima indicates that the star-like structures of the PSD plots (in Fig.\,\ref{eudistance}) have a 7-fold symmetry and might rotate with a rigid $ 360 ^\circ$ motion during one complete synchrotron oscillation period, similar to the model in \cite{Roussel2014}. This situation might change for other times, different dynamic regimes of the MBI and different experimental settings of the storage ring. Since we did not enhance the contrast of the PSD reconstruction by mean bunch profile subtraction for this correlation study, this method demonstrates how situations with even challenging signal-to-noise ratios can be handled.\\
Finally, the Euclidean distance shows a slow but sig\-ni\-fi\-cant up and down bending, particularly visible at the progression of the minima positions. We attribute this effect to a mismatch between the rotation center and the center of gravity of the PSD (rather than an asymmetry of the PSD itself), which is consistent with the observation of the synchrotron oscillation in the revolution plots \cite{Ormond} (even when the substructures are faded out, see especially Fig. \ref{experiment}).
\section{Summary and Outlook}
In this publication, we demonstrated a robust method to reconstruct the longitudinal PSD from turn-by-turn measurements of the longitudinal bunch profile using a filtered back-projection. In doing so we close an im\-por\-tant diagnostics gap for electron storage rings. Our approach is based on the  approximation of the electron bunch dynamics by a rigid rotation of the PSD during half a synchrotron oscillation period. From comparison to simulations, we provide em\-pi\-ri\-cal evidence, that the filtered back-projection performs very well, even under conditions with strong intra-bunch interactions. Applying the PSD reconstruction to experimental data, we are able to observe a typical dynamic cycle in a reproducible manner.\\ For future research, it might be interesting to analyze the filtered back-projection performance with respect to the PSD dynamics governed by the Vlasov-Fokker-Planck equation more deeply. Moreover, considering that the filtered back-projection is originally not intended to be used for PSD tomography of electron bunches, it might be possible to construct a specialized algorithm deduced from the Vlasov-Fokker-Planck equation relaxing the rigid rotation assumption.\\
Finally, our approach offers a glimpse into the physics of equilibrium/non-equilibrium ultrarelativistic systems. It therefore supports the development of advanced accelerator concepts with low energy footprints, where coherent emission from compressed, high-density  electron beams plays an import role \cite{Seletskiy2013}.  

\section{Methods}

\begin{figure*}
\includegraphics[width=\linewidth]{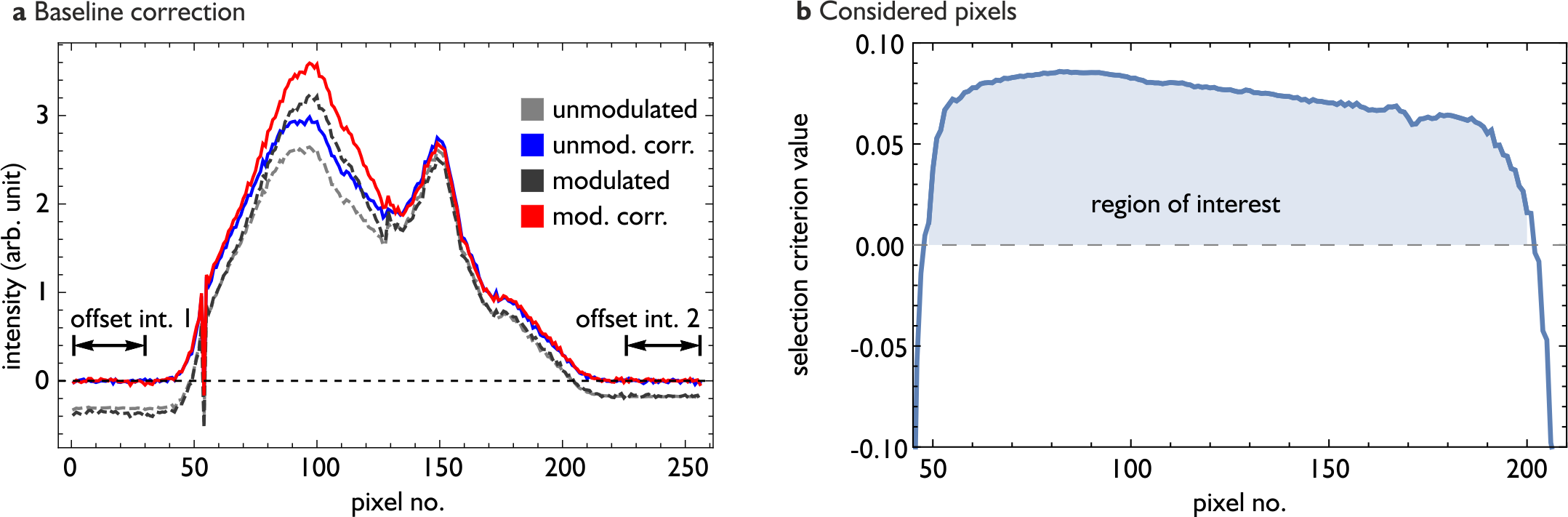}
\caption{\label{OffsetRemoval} a) Removal of the Gotthard chip dependent offset from the measured data. b) Definition of the region of interest.}
\end{figure*}
\begin{figure}
\includegraphics[width=\linewidth]{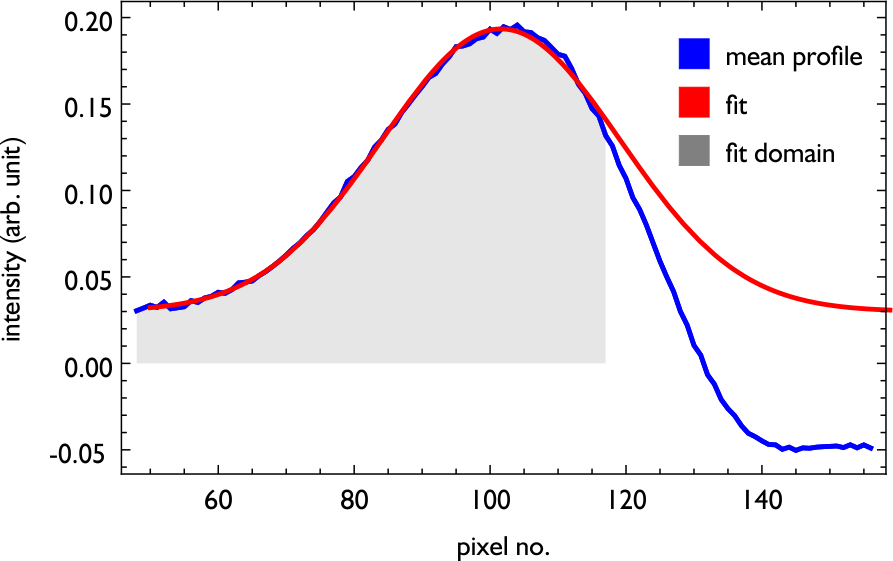}
\caption{\label{CenterpointDet} Determination of the rotation center shown for the case of first revolution plot in Fig. \ref{experiment}. Pixel no. 102 is located closest to the center point.}
\end{figure}

\subsection{Experimental parameters and setup}

The measurements were enabled through an synchronized interplay between three sophisticated instruments: an accelerator providing strongly self-interacting relativistic electron bunches, a single-shot EO sampling experiment using a synchronized laser system and a fast line array camera system measuring at MHz repetition rates.  \\
The experiments were conducted at the Karlsruhe Research Accelerator (KARA), which was operated in a dedicated short bunch mode \cite{Muller2005} to compress the bunch length to a few picoseconds. The machine parameter settings, which were similar to those in  reference \cite{Steinmann2018}, are displayed in Table \ref{storage_ring_parameters}. During the measurements, the beam energy was set to 1.3 GeV using a radio frequency voltage of 799 kV. The data of Fig.\,\ref{experiment} was taken at a bunch current of 0.85 mA in single bunch operation mode at the so-called sawtooth bursting regime (see \cite{Steinmann2018}). \\
For the EO measurement, we used chirped laser pulses from a custom-build regenerative amplifier \cite{Muller2012} with a central wavelength of 1050 nm, an optical power of a few mW and a spectral width of about 80 nm. The ytterbium-doped fiber oscillator of the laser system operates at 62.5 MHz and is actively synchronized to the 500 MHz of the master oscillator. A pulse-picker is then used to reduce the repetition rate of the laser system (see also \cite{Funkner2019} for further details).\\
Afterwards, the spectrum of the laser pulses was resolved by a grating and measured with a KIT-developed ultra-fast spectrometer KALYPSO \cite{Rota2018} (KArlsruhe Linear arraY detector for MHz-rePetition rate SpectrOscopy) using a 256-pixel back-illuminated silicon line array. The repetition rate of the line array readout was 2.72 MHz, so that that the electron bunch profile was detected on a turn-by-turn basis. 

\begin{table}
\caption{\label{storage_ring_parameters} Machine parameters during the experiment.}
\begin{ruledtabular}
\begin{tabular}{ll}
\hline
Parameter & Value \\
\hline
Beam energy & 1.3 GeV \\
Calculated relative energy spread & $4.7 \cdot 10^{-4}$\\
Bunch current &0.85 mA \\
Radio frequency voltage &799 kV \\
Radio frequency & 499.73 MHz\\
Revolution frequency & 2.7159 MHz \\
Synchrotron frequency & 8.28 kHz\\
Calculated momentum compaction factor & $5\cdot 10^{-4}$ \\
Circumference & 110.4 m \\
Vacuum chamber height & 32 mm
\end{tabular}
\end{ruledtabular}
\end{table}
\subsection{Details of the post-measurement data processing and the back-projection parameters}
To calculate the bunch profile, we first determine the background signal of the line array by blocking the laser beam. In the next step, we record the signal of the laser pulses propagation through the EO (GaP) crystal without the overlap with the electron bunch. To do so, the phase of the laser  synchronization system is changed, so that the laser pulse arrives before the electron bunch at the EO crystal. We subtract the background signal from the measurements with and without overlap and obtain the modulated and unmodulated signal. The result is shown in Fig.\,\ref{OffsetRemoval}\,a. \\
The gray and black curves display the unmodulated and modulated signals, respectively. The edges of the line array are not illuminated by the laser pulses. The data acquisition of the 256-pixel line array is realized by two Gotthard chips reading out 128 pixel arrays each. The negative values at the intervals pictured in Fig.\,\ref{OffsetRemoval}\,a and the kink between the pixel 128 and pixel 129 indicate that the measured intensities are reduced by a constant, distinct offset for each Gotthard chip.  For further calculations, we determined these offsets by averaging the intensity for all measurements in the region indicated by the intervals in Fig.\,\ref{OffsetRemoval}\,a. Afterwards, the offsets were removed from every data set. Figure \ref{OffsetRemoval}\,a displays in red and blue the corrected values for the modulated and unmodulated signals. Especially the kink is reduced significantly for these data sets. With the corrected data, we calculated the bunch profiles according to \cite{Funkner2019} by the division between the modulated and the unmodulated signal. \\ 
In the next step we determined which pixel should be considered for further analysis. Clearly such a decision depends on the signal-to-noise ratio for the calculated charge density of the bunch profiles. Here we used the following criterion: we consider all pixels for which half of the maximum signal modulation can be distinguished from a background signal with an accuracy of 90 \%. We assume that the maximum signal is about 0.2 (determined from average bunch profiles). As described in \cite{Funkner2019}, let $\sigma_{\tilde{\rho},i}$ be the standard deviation for a hypothetical repeated, single-shot measurement for pixel $i$, then we consider all pixel for which:
\begin{equation}
    \text{selection criterion value} : = 0.1-1.645 \sigma_{\tilde{\rho},i} > 0 \label{pixelcriterion}
\end{equation}
is fulfilled. Figure \ref{OffsetRemoval}\,b displays the dependence of $\sigma_{\tilde{\rho}}$ on the pixel number. From this procedure the pixels numbers between 48 and 201 are selected. While \eqref{pixelcriterion} provides an objective and reproducible measure, which data should be considered for further evaluation,  the reconstruction of the PSD needs further refinement. \\
For the filtered back-projection it is assumed that the PSD rotates around a center point, which is located in the middle of the sinogram. Thus, we have to determine at which pixel the center point of rotation is located and then further cut the considered data range symmetrically around this center point. In order to calculate this center point, we average all bunch profiles in a section  for which the reconstruction is considered (see Fig.\,\ref{CenterpointDet}). We then fit a normal distribution to the bunch profile considering pixels from 48 to 114, because later the profile data is overlapped the wake field of the electron bunch \cite{SchoenfeldtBlomleyBruendermann2017_1000071492}, which results in a distortion of the line shape. Figure \ref{CenterpointDet} visualizes this calculation for the first section in Fig. \ref{experiment}a. The mean of the normal distribution is closest to pixel 102. For the other sections in Fig. \ref{experiment}b, we obtain similar fits and with pixel 102 and pixel 101 similar corresponding means. Finally, we chose the largest interval for which the center rotation point is in the middle and which is part of the interval from pixel 48 and 201. Thus, for the first two revolution plots, we used the pixels between pixel 48 and 156 and for the last revolution plot we used the pixels between 48 and 154. In general, these small differences have no influence on the data interpretation at the current signal-to-noise ratio, i.e. it would be also possible to use the same interval for all reconstructions. 
\subsection{Validation of the back-projection method with simulations}

In general, our motivating assumption that the dynamics of an electron bunch can be approximated by a rigid rotation is a strong simplification. In fact, static deformation~\cite{haissinski1973, ng_2006} as well as the dynamic formation of substructures~\cite{Krinsky:1985sq} is caused by nonlinearities in the potential.\\
Thus, to validate the reconstruction empirically under the conditions present during the experiment, we simulate the PSD dynamics by numerically solving the Vlasov-Fokker-Planck equation using the Inovesa software program. Our group already demonstrated that the software can precisely predict the bunch current depending bursting behavior of electron bunches
over long time scales \cite{Schonfeldt2017}.\\
The PSDs shown in Fig. \ref{reconstruction} were simulated using Ino\-ve\-sa v1.1.0 (\href{https://dx.doi.org/10.5281/zenodo.3466767}{doi:10.5281/zenodo.3466767}). This version implements the phase noise of the radio frequency cavities to excite the synchrotron oscillation, which is ty\-pi\-cal\-ly not part of simulations \cite{Schonfeldt2018a}.  The resulting potential is modified by the wake potential due to self-interaction with the CSR of the bunch and calculated according to the parallel plates shielding model~\cite{murphy1997}. The simulations were carried out, closely to the experimental conditions, with a bunch current of 0.85 mA at the sawtooth bursting regime. The list of physical parameters is displayed in Table ~\ref{inovesa_parameters}.
\begin{table}
\caption{\label{inovesa_parameters} Parameter settings used for the Inovesa simulation of the PSD dynamics.}
\begin{ruledtabular}
\begin{tabular}{ll}
\hline
Parameter & Value \\
\hline
Beam energy & 1.285 GeV \\
Natural relative energy spread & $4.7 \cdot 10^{-4}$\\
Bunch current & 0.85 mA \\
Radio frequency voltage & 799.257 kV \\
Revolution frequency & 2.71592 MHz \\
Synchrotron frequency & 8.256 kHz\\
Damping time & 10.4 ms \\
Harmonic number & 184 \\
Vacuum chamber height & 32 mm\\
\end{tabular}
\end{ruledtabular}
\end{table}

\section{Acknowledgment}

S.F. and B.K. were supported by BMBF contract No. 05K16VKA. E.B. and G.N. acknowledge support by the Helmholtz President’s strategic fund IVF ``Plasma accelerators''. The authors thank N. Hiller, B. Härer, T. Boltz, M. Brosi and P. Schreiber for fruitful discussions. We further thank J. Schmid and O. Speck for their technical support.

\bibliography{EO_phase_space_tomography_02}

\begin{thebibliography}{50}%
\makeatletter
\providecommand \@ifxundefined [1]{%
 \@ifx{#1\undefined}
}%
\providecommand \@ifnum [1]{%
 \ifnum #1\expandafter \@firstoftwo
 \else \expandafter \@secondoftwo
 \fi
}%
\providecommand \@ifx [1]{%
 \ifx #1\expandafter \@firstoftwo
 \else \expandafter \@secondoftwo
 \fi
}%
\providecommand \natexlab [1]{#1}%
\providecommand \enquote  [1]{``#1''}%
\providecommand \bibnamefont  [1]{#1}%
\providecommand \bibfnamefont [1]{#1}%
\providecommand \citenamefont [1]{#1}%
\providecommand \href@noop [0]{\@secondoftwo}%
\providecommand \href [0]{\begingroup \@sanitize@url \@href}%
\providecommand \@href[1]{\@@startlink{#1}\@@href}%
\providecommand \@@href[1]{\endgroup#1\@@endlink}%
\providecommand \@sanitize@url [0]{\catcode `\\12\catcode `\$12\catcode
  `\&12\catcode `\#12\catcode `\^12\catcode `\_12\catcode `\%12\relax}%
\providecommand \@@startlink[1]{}%
\providecommand \@@endlink[0]{}%
\providecommand \url  [0]{\begingroup\@sanitize@url \@url }%
\providecommand \@url [1]{\endgroup\@href {#1}{\urlprefix }}%
\providecommand \urlprefix  [0]{URL }%
\providecommand \Eprint [0]{\href }%
\providecommand \doibase [0]{http://dx.doi.org/}%
\providecommand \selectlanguage [0]{\@gobble}%
\providecommand \bibinfo  [0]{\@secondoftwo}%
\providecommand \bibfield  [0]{\@secondoftwo}%
\providecommand \translation [1]{[#1]}%
\providecommand \BibitemOpen [0]{}%
\providecommand \bibitemStop [0]{}%
\providecommand \bibitemNoStop [0]{.\EOS\space}%
\providecommand \EOS [0]{\spacefactor3000\relax}%
\providecommand \BibitemShut  [1]{\csname bibitem#1\endcsname}%
\let\auto@bib@innerbib\@empty
\bibitem [{\citenamefont {Arnold}\ \emph {et~al.}(2004)\citenamefont {Arnold},
  \citenamefont {Carrillo}, \citenamefont {Desvillettes}, \citenamefont
  {Dolbeault}, \citenamefont {J{\"{u}}ngel}, \citenamefont {Lederman},
  \citenamefont {Markowich}, \citenamefont {Toscani},\ and\ \citenamefont
  {Villani}}]{Arnold2004}%
  \BibitemOpen
  \bibfield  {author} {\bibinfo {author} {\bibfnamefont {A.}~\bibnamefont
  {Arnold}}, \bibinfo {author} {\bibfnamefont {J.~A.}\ \bibnamefont
  {Carrillo}}, \bibinfo {author} {\bibfnamefont {L.}~\bibnamefont
  {Desvillettes}}, \bibinfo {author} {\bibfnamefont {J.}~\bibnamefont
  {Dolbeault}}, \bibinfo {author} {\bibfnamefont {A.}~\bibnamefont
  {J{\"{u}}ngel}}, \bibinfo {author} {\bibfnamefont {C.}~\bibnamefont
  {Lederman}}, \bibinfo {author} {\bibfnamefont {P.~A.}\ \bibnamefont
  {Markowich}}, \bibinfo {author} {\bibfnamefont {G.}~\bibnamefont {Toscani}},
  \ and\ \bibinfo {author} {\bibfnamefont {C.}~\bibnamefont {Villani}},\ }\href
  {\doibase 10.1007/s00605-004-0239-2} {\bibfield  {journal} {\bibinfo
  {journal} {Monatshefte f{\"{u}}r Mathematik}\ }\textbf {\bibinfo {volume}
  {142}},\ \bibinfo {pages} {35} (\bibinfo {year} {2004})}\BibitemShut
  {NoStop}%
\bibitem [{\citenamefont {Alder}\ and\ \citenamefont
  {Wainwright}(1959)}]{Alder1959}%
  \BibitemOpen
  \bibfield  {author} {\bibinfo {author} {\bibfnamefont {B.~J.}\ \bibnamefont
  {Alder}}\ and\ \bibinfo {author} {\bibfnamefont {T.~E.}\ \bibnamefont
  {Wainwright}},\ }\href {\doibase 10.1063/1.1730376} {\bibfield  {journal}
  {\bibinfo  {journal} {The Journal of Chemical Physics}\ }\textbf {\bibinfo
  {volume} {31}},\ \bibinfo {pages} {459} (\bibinfo {year} {1959})}\BibitemShut
  {NoStop}%
\bibitem [{\citenamefont {Helbing}(2001)}]{Helbing2001}%
  \BibitemOpen
  \bibfield  {author} {\bibinfo {author} {\bibfnamefont {D.}~\bibnamefont
  {Helbing}},\ }\href {\doibase 10.1103/RevModPhys.73.1067} {\bibfield
  {journal} {\bibinfo  {journal} {Reviews of Modern Physics}\ }\textbf
  {\bibinfo {volume} {73}},\ \bibinfo {pages} {1067} (\bibinfo {year}
  {2001})}\BibitemShut {NoStop}%
\bibitem [{\citenamefont {Negele}\ and\ \citenamefont
  {Orland}(2018)}]{Negele2018}%
  \BibitemOpen
  \bibfield  {author} {\bibinfo {author} {\bibfnamefont {J.~W.}\ \bibnamefont
  {Negele}}\ and\ \bibinfo {author} {\bibfnamefont {H.}~\bibnamefont
  {Orland}},\ }\href {\doibase 10.1201/9780429497926} {\emph {\bibinfo {title}
  {{Quantum Many-Particle Systems}}}},\ \bibinfo {edition} {1st}\ ed.\
  (\bibinfo  {publisher} {CRC Press},\ \bibinfo {address} {Boca Raton},\
  \bibinfo {year} {2018})\ p.\ \bibinfo {pages} {476}\BibitemShut {NoStop}%
\bibitem [{\citenamefont {Schwinger}(1996)}]{Schwinger1996}%
  \BibitemOpen
  \bibfield  {author} {\bibinfo {author} {\bibfnamefont {J.}~\bibnamefont
  {Schwinger}},\ }\href {\doibase 10.2172/1195620} {\emph {\bibinfo {title}
  {{On radiation by electrons in a betatron: Transcription of a paper by J.
  Schwinger, 1945}}}},\ \bibinfo {type} {Tech. Rep.}\ (\bibinfo  {institution}
  {Lawrence Berkeley National Laboratory (LBNL)},\ \bibinfo {address}
  {Berkeley, CA (United States)},\ \bibinfo {year} {1996})\BibitemShut
  {NoStop}%
\bibitem [{\citenamefont {Schiff}(1946)}]{Schiff1946}%
  \BibitemOpen
  \bibfield  {author} {\bibinfo {author} {\bibfnamefont {L.~I.}\ \bibnamefont
  {Schiff}},\ }\href {\doibase 10.1063/1.1770395} {\bibfield  {journal}
  {\bibinfo  {journal} {Review of Scientific Instruments}\ }\textbf {\bibinfo
  {volume} {17}},\ \bibinfo {pages} {6} (\bibinfo {year} {1946})}\BibitemShut
  {NoStop}%
\bibitem [{\citenamefont {Billinghurst}\ \emph {et~al.}(2016)\citenamefont
  {Billinghurst}, \citenamefont {Bergstrom}, \citenamefont {Baribeau},
  \citenamefont {Batten}, \citenamefont {May}, \citenamefont {Vogt},\ and\
  \citenamefont {Wurtz}}]{PhysRevAccelBeams.19.020704}%
  \BibitemOpen
  \bibfield  {author} {\bibinfo {author} {\bibfnamefont {B.~E.}\ \bibnamefont
  {Billinghurst}}, \bibinfo {author} {\bibfnamefont {J.~C.}\ \bibnamefont
  {Bergstrom}}, \bibinfo {author} {\bibfnamefont {C.}~\bibnamefont {Baribeau}},
  \bibinfo {author} {\bibfnamefont {T.}~\bibnamefont {Batten}}, \bibinfo
  {author} {\bibfnamefont {T.~E.}\ \bibnamefont {May}}, \bibinfo {author}
  {\bibfnamefont {J.~M.}\ \bibnamefont {Vogt}}, \ and\ \bibinfo {author}
  {\bibfnamefont {W.~A.}\ \bibnamefont {Wurtz}},\ }\href {\doibase
  10.1103/PhysRevAccelBeams.19.020704} {\bibfield  {journal} {\bibinfo
  {journal} {Phys. Rev. Accel. Beams}\ }\textbf {\bibinfo {volume} {19}},\
  \bibinfo {pages} {20704} (\bibinfo {year} {2016})}\BibitemShut {NoStop}%
\bibitem [{\citenamefont {Tammaro}\ \emph {et~al.}(2015)\citenamefont
  {Tammaro}, \citenamefont {Pirali}, \citenamefont {Roy}, \citenamefont
  {Lampin}, \citenamefont {Ducournau}, \citenamefont {Cuisset}, \citenamefont
  {Hindle},\ and\ \citenamefont {Mouret}}]{Tammaro2015}%
  \BibitemOpen
  \bibfield  {author} {\bibinfo {author} {\bibfnamefont {S.}~\bibnamefont
  {Tammaro}}, \bibinfo {author} {\bibfnamefont {O.}~\bibnamefont {Pirali}},
  \bibinfo {author} {\bibfnamefont {P.}~\bibnamefont {Roy}}, \bibinfo {author}
  {\bibfnamefont {J.-F.}\ \bibnamefont {Lampin}}, \bibinfo {author}
  {\bibfnamefont {G.}~\bibnamefont {Ducournau}}, \bibinfo {author}
  {\bibfnamefont {A.}~\bibnamefont {Cuisset}}, \bibinfo {author} {\bibfnamefont
  {F.}~\bibnamefont {Hindle}}, \ and\ \bibinfo {author} {\bibfnamefont
  {G.}~\bibnamefont {Mouret}},\ }\href {\doibase 10.1038/ncomms8733} {\bibfield
   {journal} {\bibinfo  {journal} {Nature Communications}\ }\textbf {\bibinfo
  {volume} {6}},\ \bibinfo {pages} {7733} (\bibinfo {year} {2015})}\BibitemShut
  {NoStop}%
\bibitem [{\citenamefont {Stupakov}\ and\ \citenamefont
  {Heifets}(2002)}]{Stupakov2002}%
  \BibitemOpen
  \bibfield  {author} {\bibinfo {author} {\bibfnamefont {G.}~\bibnamefont
  {Stupakov}}\ and\ \bibinfo {author} {\bibfnamefont {S.}~\bibnamefont
  {Heifets}},\ }\href {\doibase 10.1103/PhysRevSTAB.5.054402} {\bibfield
  {journal} {\bibinfo  {journal} {Physical Review Special Topics - Accelerators
  and Beams}\ }\textbf {\bibinfo {volume} {5}},\ \bibinfo {pages} {054402}
  (\bibinfo {year} {2002})}\BibitemShut {NoStop}%
\bibitem [{\citenamefont {Kaufmann}\ and\ \citenamefont
  {Raulin}(2006)}]{Kaufmann2006}%
  \BibitemOpen
  \bibfield  {author} {\bibinfo {author} {\bibfnamefont {P.}~\bibnamefont
  {Kaufmann}}\ and\ \bibinfo {author} {\bibfnamefont {J.-P.}\ \bibnamefont
  {Raulin}},\ }\href {\doibase 10.1063/1.2244526} {\bibfield  {journal}
  {\bibinfo  {journal} {Physics of Plasmas}\ }\textbf {\bibinfo {volume}
  {13}},\ \bibinfo {pages} {070701} (\bibinfo {year} {2006})}\BibitemShut
  {NoStop}%
\bibitem [{\citenamefont {Cross}\ and\ \citenamefont
  {Hohenberg}(1993)}]{Cross1993}%
  \BibitemOpen
  \bibfield  {author} {\bibinfo {author} {\bibfnamefont {M.~C.}\ \bibnamefont
  {Cross}}\ and\ \bibinfo {author} {\bibfnamefont {P.~C.}\ \bibnamefont
  {Hohenberg}},\ }\href {\doibase 10.1103/RevModPhys.65.851} {\bibfield
  {journal} {\bibinfo  {journal} {Reviews of Modern Physics}\ }\textbf
  {\bibinfo {volume} {65}},\ \bibinfo {pages} {851} (\bibinfo {year}
  {1993})}\BibitemShut {NoStop}%
\bibitem [{\citenamefont {Evain}\ \emph {et~al.}(2019)\citenamefont {Evain},
  \citenamefont {Szwaj}, \citenamefont {Roussel}, \citenamefont {Rodriguez},
  \citenamefont {{Le Parquier}}, \citenamefont {Tordeux}, \citenamefont
  {Ribeiro}, \citenamefont {Labat}, \citenamefont {Hubert}, \citenamefont
  {Brubach}, \citenamefont {Roy},\ and\ \citenamefont {Bielawski}}]{Evain2019}%
  \BibitemOpen
  \bibfield  {author} {\bibinfo {author} {\bibfnamefont {C.}~\bibnamefont
  {Evain}}, \bibinfo {author} {\bibfnamefont {C.}~\bibnamefont {Szwaj}},
  \bibinfo {author} {\bibfnamefont {E.}~\bibnamefont {Roussel}}, \bibinfo
  {author} {\bibfnamefont {J.}~\bibnamefont {Rodriguez}}, \bibinfo {author}
  {\bibfnamefont {M.}~\bibnamefont {{Le Parquier}}}, \bibinfo {author}
  {\bibfnamefont {M.-A.}\ \bibnamefont {Tordeux}}, \bibinfo {author}
  {\bibfnamefont {F.}~\bibnamefont {Ribeiro}}, \bibinfo {author} {\bibfnamefont
  {M.}~\bibnamefont {Labat}}, \bibinfo {author} {\bibfnamefont
  {N.}~\bibnamefont {Hubert}}, \bibinfo {author} {\bibfnamefont {J.-B.}\
  \bibnamefont {Brubach}}, \bibinfo {author} {\bibfnamefont {P.}~\bibnamefont
  {Roy}}, \ and\ \bibinfo {author} {\bibfnamefont {S.}~\bibnamefont
  {Bielawski}},\ }\href {\doibase 10.1038/s41567-019-0488-6} {\bibfield
  {journal} {\bibinfo  {journal} {Nature Physics}\ }\textbf {\bibinfo {volume}
  {15}},\ \bibinfo {pages} {635} (\bibinfo {year} {2019})}\BibitemShut
  {NoStop}%
\bibitem [{\citenamefont {Roussel}\ \emph {et~al.}(2015)\citenamefont
  {Roussel}, \citenamefont {Evain}, \citenamefont {{Le Parquier}},
  \citenamefont {Szwaj}, \citenamefont {Bielawski}, \citenamefont {Manceron},
  \citenamefont {Brubach}, \citenamefont {Tordeux}, \citenamefont {Ricaud},
  \citenamefont {Cassinari}, \citenamefont {Labat}, \citenamefont {Couprie},\
  and\ \citenamefont {Roy}}]{Roussel2015a}%
  \BibitemOpen
  \bibfield  {author} {\bibinfo {author} {\bibfnamefont {E.}~\bibnamefont
  {Roussel}}, \bibinfo {author} {\bibfnamefont {C.}~\bibnamefont {Evain}},
  \bibinfo {author} {\bibfnamefont {M.}~\bibnamefont {{Le Parquier}}}, \bibinfo
  {author} {\bibfnamefont {C.}~\bibnamefont {Szwaj}}, \bibinfo {author}
  {\bibfnamefont {S.}~\bibnamefont {Bielawski}}, \bibinfo {author}
  {\bibfnamefont {L.}~\bibnamefont {Manceron}}, \bibinfo {author}
  {\bibfnamefont {J.-B.}\ \bibnamefont {Brubach}}, \bibinfo {author}
  {\bibfnamefont {M.-A.}\ \bibnamefont {Tordeux}}, \bibinfo {author}
  {\bibfnamefont {J.-P.}\ \bibnamefont {Ricaud}}, \bibinfo {author}
  {\bibfnamefont {L.}~\bibnamefont {Cassinari}}, \bibinfo {author}
  {\bibfnamefont {M.}~\bibnamefont {Labat}}, \bibinfo {author} {\bibfnamefont
  {M.-E.}\ \bibnamefont {Couprie}}, \ and\ \bibinfo {author} {\bibfnamefont
  {P.}~\bibnamefont {Roy}},\ }\href {\doibase 10.1038/srep10330} {\bibfield
  {journal} {\bibinfo  {journal} {Scientific Reports}\ }\textbf {\bibinfo
  {volume} {5}},\ \bibinfo {pages} {10330} (\bibinfo {year}
  {2015})}\BibitemShut {NoStop}%
\bibitem [{\citenamefont {W{\"{u}}stefeld}\ \emph {et~al.}(2010)\citenamefont
  {W{\"{u}}stefeld}, \citenamefont {Feikes}, \citenamefont {Hatrott},
  \citenamefont {Ries}, \citenamefont {Hoehl}, \citenamefont {Klein},
  \citenamefont {M{\"{u}}ller}, \citenamefont {Serdyukov},\ and\ \citenamefont
  {Ulm}}]{wustefeld2010coherent}%
  \BibitemOpen
  \bibfield  {author} {\bibinfo {author} {\bibfnamefont {G.}~\bibnamefont
  {W{\"{u}}stefeld}}, \bibinfo {author} {\bibfnamefont {J.}~\bibnamefont
  {Feikes}}, \bibinfo {author} {\bibfnamefont {M.~v.}\ \bibnamefont {Hatrott}},
  \bibinfo {author} {\bibfnamefont {M.}~\bibnamefont {Ries}}, \bibinfo {author}
  {\bibfnamefont {A.}~\bibnamefont {Hoehl}}, \bibinfo {author} {\bibfnamefont
  {R.}~\bibnamefont {Klein}}, \bibinfo {author} {\bibfnamefont
  {R.}~\bibnamefont {M{\"{u}}ller}}, \bibinfo {author} {\bibfnamefont
  {A.}~\bibnamefont {Serdyukov}}, \ and\ \bibinfo {author} {\bibfnamefont
  {G.}~\bibnamefont {Ulm}},\ }in\ \href@noop {} {\emph {\bibinfo {booktitle}
  {Proceedings of the first International Particle Accelerator Conference,
  Kyoto, Japan}}}\ (\bibinfo  {publisher} {JACoW},\ \bibinfo {year} {2010})\
  pp.\ \bibinfo {pages} {2508--2510}\BibitemShut {NoStop}%
\bibitem [{\citenamefont {Abo-Bakr}\ \emph {et~al.}(2003)\citenamefont
  {Abo-Bakr}, \citenamefont {Feikes}, \citenamefont {Holldack}, \citenamefont
  {Kuske},\ and\ \citenamefont {W{\"{u}}stefeld}}]{Abo-Bakr2003}%
  \BibitemOpen
  \bibfield  {author} {\bibinfo {author} {\bibfnamefont {M.}~\bibnamefont
  {Abo-Bakr}}, \bibinfo {author} {\bibfnamefont {J.}~\bibnamefont {Feikes}},
  \bibinfo {author} {\bibfnamefont {K.}~\bibnamefont {Holldack}}, \bibinfo
  {author} {\bibfnamefont {P.}~\bibnamefont {Kuske}}, \ and\ \bibinfo {author}
  {\bibfnamefont {G.}~\bibnamefont {W{\"{u}}stefeld}},\ }in\ \href {\doibase
  10.1109/PAC.2003.1289801} {\emph {\bibinfo {booktitle} {Proceedings of the
  2003 Bipolar/BiCMOS Circuits and Technology Meeting}}},\ Vol.~\bibinfo
  {volume} {5}\ (\bibinfo  {publisher} {IEEE},\ \bibinfo {year} {2003})\ pp.\
  \bibinfo {pages} {3023--3025}\BibitemShut {NoStop}%
\bibitem [{\citenamefont {Shields}\ \emph {et~al.}(2012)\citenamefont
  {Shields}, \citenamefont {Bartolini}, \citenamefont {Boorman}, \citenamefont
  {Karataev}, \citenamefont {Lyapin}, \citenamefont {Puntree},\ and\
  \citenamefont {Rehm}}]{Shields2012}%
  \BibitemOpen
  \bibfield  {author} {\bibinfo {author} {\bibfnamefont {W.}~\bibnamefont
  {Shields}}, \bibinfo {author} {\bibfnamefont {R.}~\bibnamefont {Bartolini}},
  \bibinfo {author} {\bibfnamefont {G.}~\bibnamefont {Boorman}}, \bibinfo
  {author} {\bibfnamefont {P.}~\bibnamefont {Karataev}}, \bibinfo {author}
  {\bibfnamefont {A.}~\bibnamefont {Lyapin}}, \bibinfo {author} {\bibfnamefont
  {J.}~\bibnamefont {Puntree}}, \ and\ \bibinfo {author} {\bibfnamefont
  {G.}~\bibnamefont {Rehm}},\ }\href {\doibase 10.1088/1742-6596/357/1/012037}
  {\bibfield  {journal} {\bibinfo  {journal} {Journal of Physics: Conference
  Series}\ }\textbf {\bibinfo {volume} {357}},\ \bibinfo {pages} {012037}
  (\bibinfo {year} {2012})}\BibitemShut {NoStop}%
\bibitem [{\citenamefont {Warnock}\ and\ \citenamefont
  {Ellison}(2000)}]{warnock2000general}%
  \BibitemOpen
  \bibfield  {author} {\bibinfo {author} {\bibfnamefont {R.~L.}\ \bibnamefont
  {Warnock}}\ and\ \bibinfo {author} {\bibfnamefont {J.~A.}\ \bibnamefont
  {Ellison}},\ }in\ \href@noop {} {\emph {\bibinfo {booktitle} {Proc. 2nd ICFA
  Advanced Workshop on The Physics of High Brightness Beams (University of
  California, LA, 1999)}}}\ (\bibinfo {year} {2000})\ pp.\ \bibinfo {pages}
  {322--348}\BibitemShut {NoStop}%
\bibitem [{\citenamefont {Bane}\ \emph {et~al.}(2010)\citenamefont {Bane},
  \citenamefont {Cai},\ and\ \citenamefont {Stupakov}}]{Bane2010}%
  \BibitemOpen
  \bibfield  {author} {\bibinfo {author} {\bibfnamefont {K.~L.~F.}\
  \bibnamefont {Bane}}, \bibinfo {author} {\bibfnamefont {Y.}~\bibnamefont
  {Cai}}, \ and\ \bibinfo {author} {\bibfnamefont {G.}~\bibnamefont
  {Stupakov}},\ }\href {\doibase 10.1103/PhysRevSTAB.13.104402} {\bibfield
  {journal} {\bibinfo  {journal} {Physical Review Special Topics - Accelerators
  and Beams}\ }\textbf {\bibinfo {volume} {13}},\ \bibinfo {pages} {104402}
  (\bibinfo {year} {2010})}\BibitemShut {NoStop}%
\bibitem [{\citenamefont {Evain}\ \emph {et~al.}(2012)\citenamefont {Evain},
  \citenamefont {Barros}, \citenamefont {Loulergue}, \citenamefont {Tordeux},
  \citenamefont {Nagaoka}, \citenamefont {Labat}, \citenamefont {Cassinari},
  \citenamefont {Creff}, \citenamefont {Manceron}, \citenamefont {Brubach},
  \citenamefont {Roy},\ and\ \citenamefont {Couprie}}]{Evain2012}%
  \BibitemOpen
  \bibfield  {author} {\bibinfo {author} {\bibfnamefont {C.}~\bibnamefont
  {Evain}}, \bibinfo {author} {\bibfnamefont {J.}~\bibnamefont {Barros}},
  \bibinfo {author} {\bibfnamefont {A.}~\bibnamefont {Loulergue}}, \bibinfo
  {author} {\bibfnamefont {M.~A.}\ \bibnamefont {Tordeux}}, \bibinfo {author}
  {\bibfnamefont {R.}~\bibnamefont {Nagaoka}}, \bibinfo {author} {\bibfnamefont
  {M.}~\bibnamefont {Labat}}, \bibinfo {author} {\bibfnamefont
  {L.}~\bibnamefont {Cassinari}}, \bibinfo {author} {\bibfnamefont
  {G.}~\bibnamefont {Creff}}, \bibinfo {author} {\bibfnamefont
  {L.}~\bibnamefont {Manceron}}, \bibinfo {author} {\bibfnamefont {J.~B.}\
  \bibnamefont {Brubach}}, \bibinfo {author} {\bibfnamefont {P.}~\bibnamefont
  {Roy}}, \ and\ \bibinfo {author} {\bibfnamefont {M.~E.}\ \bibnamefont
  {Couprie}},\ }\href {\doibase 10.1209/0295-5075/98/40006} {\bibfield
  {journal} {\bibinfo  {journal} {EPL (Europhysics Letters)}\ }\textbf
  {\bibinfo {volume} {98}},\ \bibinfo {pages} {40006} (\bibinfo {year}
  {2012})}\BibitemShut {NoStop}%
\bibitem [{\citenamefont {Sch{\"{o}}nfeldt}\ \emph
  {et~al.}(2017{\natexlab{a}})\citenamefont {Sch{\"{o}}nfeldt}, \citenamefont
  {Brosi}, \citenamefont {Schwarz}, \citenamefont {Steinmann},\ and\
  \citenamefont {M{\"{u}}ller}}]{Schonfeldt2017}%
  \BibitemOpen
  \bibfield  {author} {\bibinfo {author} {\bibfnamefont {P.}~\bibnamefont
  {Sch{\"{o}}nfeldt}}, \bibinfo {author} {\bibfnamefont {M.}~\bibnamefont
  {Brosi}}, \bibinfo {author} {\bibfnamefont {M.}~\bibnamefont {Schwarz}},
  \bibinfo {author} {\bibfnamefont {J.~L.}\ \bibnamefont {Steinmann}}, \ and\
  \bibinfo {author} {\bibfnamefont {A.-S.}\ \bibnamefont {M{\"{u}}ller}},\
  }\href {\doibase 10.1103/PhysRevAccelBeams.20.030704} {\bibfield  {journal}
  {\bibinfo  {journal} {Physical Review Accelerators and Beams}\ }\textbf
  {\bibinfo {volume} {20}},\ \bibinfo {pages} {030704} (\bibinfo {year}
  {2017}{\natexlab{a}})}\BibitemShut {NoStop}%
\bibitem [{\citenamefont {Brosi}\ \emph {et~al.}(2019)\citenamefont {Brosi},
  \citenamefont {Steinmann}, \citenamefont {Blomley}, \citenamefont {Boltz},
  \citenamefont {Br{\"{u}}ndermann}, \citenamefont {Gethmann}, \citenamefont
  {Kehrer}, \citenamefont {Mathis}, \citenamefont {Papash}, \citenamefont
  {Schedler}, \citenamefont {Sch{\"{o}}nfeldt}, \citenamefont {Schreiber},
  \citenamefont {Schuh}, \citenamefont {Schwarz}, \citenamefont {M{\"{u}}ller},
  \citenamefont {Caselle}, \citenamefont {Rota}, \citenamefont {Weber},\ and\
  \citenamefont {Kuske}}]{Brosi2019}%
  \BibitemOpen
  \bibfield  {author} {\bibinfo {author} {\bibfnamefont {M.}~\bibnamefont
  {Brosi}}, \bibinfo {author} {\bibfnamefont {J.~L.}\ \bibnamefont
  {Steinmann}}, \bibinfo {author} {\bibfnamefont {E.}~\bibnamefont {Blomley}},
  \bibinfo {author} {\bibfnamefont {T.}~\bibnamefont {Boltz}}, \bibinfo
  {author} {\bibfnamefont {E.}~\bibnamefont {Br{\"{u}}ndermann}}, \bibinfo
  {author} {\bibfnamefont {J.}~\bibnamefont {Gethmann}}, \bibinfo {author}
  {\bibfnamefont {B.}~\bibnamefont {Kehrer}}, \bibinfo {author} {\bibfnamefont
  {Y.-L.}\ \bibnamefont {Mathis}}, \bibinfo {author} {\bibfnamefont
  {A.}~\bibnamefont {Papash}}, \bibinfo {author} {\bibfnamefont
  {M.}~\bibnamefont {Schedler}}, \bibinfo {author} {\bibfnamefont
  {P.}~\bibnamefont {Sch{\"{o}}nfeldt}}, \bibinfo {author} {\bibfnamefont
  {P.}~\bibnamefont {Schreiber}}, \bibinfo {author} {\bibfnamefont
  {M.}~\bibnamefont {Schuh}}, \bibinfo {author} {\bibfnamefont
  {M.}~\bibnamefont {Schwarz}}, \bibinfo {author} {\bibfnamefont {A.-S.}\
  \bibnamefont {M{\"{u}}ller}}, \bibinfo {author} {\bibfnamefont
  {M.}~\bibnamefont {Caselle}}, \bibinfo {author} {\bibfnamefont
  {L.}~\bibnamefont {Rota}}, \bibinfo {author} {\bibfnamefont {M.}~\bibnamefont
  {Weber}}, \ and\ \bibinfo {author} {\bibfnamefont {P.}~\bibnamefont
  {Kuske}},\ }\href {\doibase 10.1103/PhysRevAccelBeams.22.020701} {\bibfield
  {journal} {\bibinfo  {journal} {Physical Review Accelerators and Beams}\
  }\textbf {\bibinfo {volume} {22}},\ \bibinfo {pages} {020701} (\bibinfo
  {year} {2019})}\BibitemShut {NoStop}%
\bibitem [{\citenamefont {Carr}\ \emph {et~al.}(2001)\citenamefont {Carr},
  \citenamefont {Kramer}, \citenamefont {Murphy}, \citenamefont {Lobo},\ and\
  \citenamefont {Tanner}}]{CARR2001387}%
  \BibitemOpen
  \bibfield  {author} {\bibinfo {author} {\bibfnamefont {G.~L.}\ \bibnamefont
  {Carr}}, \bibinfo {author} {\bibfnamefont {S.~L.}\ \bibnamefont {Kramer}},
  \bibinfo {author} {\bibfnamefont {J.~B.}\ \bibnamefont {Murphy}}, \bibinfo
  {author} {\bibfnamefont {R.}~\bibnamefont {Lobo}}, \ and\ \bibinfo {author}
  {\bibfnamefont {D.~B.}\ \bibnamefont {Tanner}},\ }\href {\doibase
  https://doi.org/10.1016/S0168-9002(01)00521-6} {\bibfield  {journal}
  {\bibinfo  {journal} {Nuclear Instruments and Methods in Physics Research
  Section A: Accelerators, Spectrometers, Detectors and Associated Equipment}\
  }\textbf {\bibinfo {volume} {463}},\ \bibinfo {pages} {387} (\bibinfo {year}
  {2001})}\BibitemShut {NoStop}%
\bibitem [{\citenamefont {Byrd}\ \emph {et~al.}(2002)\citenamefont {Byrd},
  \citenamefont {Leemans}, \citenamefont {Loftsdottir}, \citenamefont
  {Marcelis}, \citenamefont {Martin}, \citenamefont {McKinney}, \citenamefont
  {Sannibale}, \citenamefont {Scarvie},\ and\ \citenamefont
  {Steier}}]{Byrd2002ObservationOB}%
  \BibitemOpen
  \bibfield  {author} {\bibinfo {author} {\bibfnamefont {J.~M.}\ \bibnamefont
  {Byrd}}, \bibinfo {author} {\bibfnamefont {W.~P.}\ \bibnamefont {Leemans}},
  \bibinfo {author} {\bibfnamefont {A.}~\bibnamefont {Loftsdottir}}, \bibinfo
  {author} {\bibfnamefont {B.}~\bibnamefont {Marcelis}}, \bibinfo {author}
  {\bibfnamefont {M.~C.}\ \bibnamefont {Martin}}, \bibinfo {author}
  {\bibfnamefont {W.~R.}\ \bibnamefont {McKinney}}, \bibinfo {author}
  {\bibfnamefont {F.}~\bibnamefont {Sannibale}}, \bibinfo {author}
  {\bibfnamefont {T.}~\bibnamefont {Scarvie}}, \ and\ \bibinfo {author}
  {\bibfnamefont {C.}~\bibnamefont {Steier}},\ }\href {\doibase
  10.1103/PhysRevLett.89.224801} {\bibfield  {journal} {\bibinfo  {journal}
  {Physical Review Letters}\ }\textbf {\bibinfo {volume} {89}},\ \bibinfo
  {pages} {224801} (\bibinfo {year} {2002})}\BibitemShut {NoStop}%
\bibitem [{\citenamefont {Karantzoulis}\ \emph {et~al.}(2010)\citenamefont
  {Karantzoulis}, \citenamefont {Penco}, \citenamefont {Perucchi},\ and\
  \citenamefont {Lupi}}]{Karantzoulis2010}%
  \BibitemOpen
  \bibfield  {author} {\bibinfo {author} {\bibfnamefont {E.}~\bibnamefont
  {Karantzoulis}}, \bibinfo {author} {\bibfnamefont {G.}~\bibnamefont {Penco}},
  \bibinfo {author} {\bibfnamefont {A.}~\bibnamefont {Perucchi}}, \ and\
  \bibinfo {author} {\bibfnamefont {S.}~\bibnamefont {Lupi}},\ }\href {\doibase
  10.1016/j.infrared.2010.04.006} {\bibfield  {journal} {\bibinfo  {journal}
  {Infrared Physics {\&} Technology}\ }\textbf {\bibinfo {volume} {53}},\
  \bibinfo {pages} {300} (\bibinfo {year} {2010})}\BibitemShut {NoStop}%
\bibitem [{\citenamefont {Steinmann}\ \emph {et~al.}(2018)\citenamefont
  {Steinmann}, \citenamefont {Boltz}, \citenamefont {Brosi}, \citenamefont
  {Br{\"{u}}ndermann}, \citenamefont {Caselle}, \citenamefont {Kehrer},
  \citenamefont {Rota}, \citenamefont {Sch{\"{o}}nfeldt}, \citenamefont
  {Schuh}, \citenamefont {Siegel}, \citenamefont {Weber},\ and\ \citenamefont
  {M{\"{u}}ller}}]{Steinmann2018}%
  \BibitemOpen
  \bibfield  {author} {\bibinfo {author} {\bibfnamefont {J.~L.}\ \bibnamefont
  {Steinmann}}, \bibinfo {author} {\bibfnamefont {T.}~\bibnamefont {Boltz}},
  \bibinfo {author} {\bibfnamefont {M.}~\bibnamefont {Brosi}}, \bibinfo
  {author} {\bibfnamefont {E.}~\bibnamefont {Br{\"{u}}ndermann}}, \bibinfo
  {author} {\bibfnamefont {M.}~\bibnamefont {Caselle}}, \bibinfo {author}
  {\bibfnamefont {B.}~\bibnamefont {Kehrer}}, \bibinfo {author} {\bibfnamefont
  {L.}~\bibnamefont {Rota}}, \bibinfo {author} {\bibfnamefont {P.}~\bibnamefont
  {Sch{\"{o}}nfeldt}}, \bibinfo {author} {\bibfnamefont {M.}~\bibnamefont
  {Schuh}}, \bibinfo {author} {\bibfnamefont {M.}~\bibnamefont {Siegel}},
  \bibinfo {author} {\bibfnamefont {M.}~\bibnamefont {Weber}}, \ and\ \bibinfo
  {author} {\bibfnamefont {A.-S.}\ \bibnamefont {M{\"{u}}ller}},\ }\href
  {\doibase 10.1103/PhysRevAccelBeams.21.110705} {\bibfield  {journal}
  {\bibinfo  {journal} {Physical Review Accelerators and Beams}\ }\textbf
  {\bibinfo {volume} {21}},\ \bibinfo {pages} {110705} (\bibinfo {year}
  {2018})}\BibitemShut {NoStop}%
\bibitem [{\citenamefont {Hancock}\ \emph {et~al.}(2000)\citenamefont
  {Hancock}, \citenamefont {Lindroos},\ and\ \citenamefont
  {Koscielniak}}]{Hancock2000}%
  \BibitemOpen
  \bibfield  {author} {\bibinfo {author} {\bibfnamefont {S.}~\bibnamefont
  {Hancock}}, \bibinfo {author} {\bibfnamefont {M.}~\bibnamefont {Lindroos}}, \
  and\ \bibinfo {author} {\bibfnamefont {S.}~\bibnamefont {Koscielniak}},\
  }\href {\doibase 10.1103/PhysRevSTAB.3.124202} {\bibfield  {journal}
  {\bibinfo  {journal} {Physical Review Special Topics - Accelerators and
  Beams}\ }\textbf {\bibinfo {volume} {3}},\ \bibinfo {pages} {124202}
  (\bibinfo {year} {2000})}\BibitemShut {NoStop}%
\bibitem [{\citenamefont {Montag}\ \emph {et~al.}(2002)\citenamefont {Montag},
  \citenamefont {D'Imperio}, \citenamefont {Kewisch},\ and\ \citenamefont
  {Lee}}]{Montag2002}%
  \BibitemOpen
  \bibfield  {author} {\bibinfo {author} {\bibfnamefont {C.}~\bibnamefont
  {Montag}}, \bibinfo {author} {\bibfnamefont {N.}~\bibnamefont {D'Imperio}},
  \bibinfo {author} {\bibfnamefont {J.}~\bibnamefont {Kewisch}}, \ and\
  \bibinfo {author} {\bibfnamefont {R.}~\bibnamefont {Lee}},\ }\href {\doibase
  10.1103/PhysRevSTAB.5.082801} {\bibfield  {journal} {\bibinfo  {journal}
  {Physical Review Special Topics - Accelerators and Beams}\ }\textbf {\bibinfo
  {volume} {5}},\ \bibinfo {pages} {082801} (\bibinfo {year}
  {2002})}\BibitemShut {NoStop}%
\bibitem [{\citenamefont {Yakimenko}\ \emph {et~al.}(2003)\citenamefont
  {Yakimenko}, \citenamefont {Babzien}, \citenamefont {Ben-Zvi}, \citenamefont
  {Malone},\ and\ \citenamefont {Wang}}]{Yakimenko2003}%
  \BibitemOpen
  \bibfield  {author} {\bibinfo {author} {\bibfnamefont {V.}~\bibnamefont
  {Yakimenko}}, \bibinfo {author} {\bibfnamefont {M.}~\bibnamefont {Babzien}},
  \bibinfo {author} {\bibfnamefont {I.}~\bibnamefont {Ben-Zvi}}, \bibinfo
  {author} {\bibfnamefont {R.}~\bibnamefont {Malone}}, \ and\ \bibinfo {author}
  {\bibfnamefont {X.-J.}\ \bibnamefont {Wang}},\ }\href {\doibase
  10.1103/PhysRevSTAB.6.122801} {\bibfield  {journal} {\bibinfo  {journal}
  {Physical Review Special Topics - Accelerators and Beams}\ }\textbf {\bibinfo
  {volume} {6}},\ \bibinfo {pages} {122801} (\bibinfo {year}
  {2003})}\BibitemShut {NoStop}%
\bibitem [{\citenamefont {Hock}\ and\ \citenamefont {Wolski}(2013)}]{Hock2013}%
  \BibitemOpen
  \bibfield  {author} {\bibinfo {author} {\bibfnamefont {K.}~\bibnamefont
  {Hock}}\ and\ \bibinfo {author} {\bibfnamefont {A.}~\bibnamefont {Wolski}},\
  }\href {\doibase 10.1016/j.nima.2013.05.004} {\bibfield  {journal} {\bibinfo
  {journal} {Nuclear Instruments and Methods in Physics Research Section A:
  Accelerators, Spectrometers, Detectors and Associated Equipment}\ }\textbf
  {\bibinfo {volume} {726}},\ \bibinfo {pages} {8} (\bibinfo {year}
  {2013})}\BibitemShut {NoStop}%
\bibitem [{\citenamefont {Michelotti}(2003)}]{Michelotti2003}%
  \BibitemOpen
  \bibfield  {author} {\bibinfo {author} {\bibfnamefont {L.}~\bibnamefont
  {Michelotti}},\ }\href {\doibase 10.1103/PhysRevSTAB.6.024001} {\bibfield
  {journal} {\bibinfo  {journal} {Physical Review Special Topics - Accelerators
  and Beams}\ }\textbf {\bibinfo {volume} {6}},\ \bibinfo {pages} {024001}
  (\bibinfo {year} {2003})}\BibitemShut {NoStop}%
\bibitem [{\citenamefont {Rota}\ \emph {et~al.}(2019)\citenamefont {Rota},
  \citenamefont {Caselle}, \citenamefont {Br{\"{u}}ndermann}, \citenamefont
  {Funkner}, \citenamefont {Gerth}, \citenamefont {Kehrer}, \citenamefont
  {Mielczarek}, \citenamefont {Makowski}, \citenamefont {Mozzanica},
  \citenamefont {M{\"{u}}ller}, \citenamefont {Nasse}, \citenamefont {Niehues},
  \citenamefont {Patil}, \citenamefont {Schmitt}, \citenamefont
  {Sch{\"{o}}nfeldt}, \citenamefont {Steffen},\ and\ \citenamefont
  {Weber}}]{Rota2018}%
  \BibitemOpen
  \bibfield  {author} {\bibinfo {author} {\bibfnamefont {L.}~\bibnamefont
  {Rota}}, \bibinfo {author} {\bibfnamefont {M.}~\bibnamefont {Caselle}},
  \bibinfo {author} {\bibfnamefont {E.}~\bibnamefont {Br{\"{u}}ndermann}},
  \bibinfo {author} {\bibfnamefont {S.}~\bibnamefont {Funkner}}, \bibinfo
  {author} {\bibfnamefont {C.}~\bibnamefont {Gerth}}, \bibinfo {author}
  {\bibfnamefont {B.}~\bibnamefont {Kehrer}}, \bibinfo {author} {\bibfnamefont
  {A.}~\bibnamefont {Mielczarek}}, \bibinfo {author} {\bibfnamefont
  {D.}~\bibnamefont {Makowski}}, \bibinfo {author} {\bibfnamefont
  {A.}~\bibnamefont {Mozzanica}}, \bibinfo {author} {\bibfnamefont {A.-S.}\
  \bibnamefont {M{\"{u}}ller}}, \bibinfo {author} {\bibfnamefont
  {M.}~\bibnamefont {Nasse}}, \bibinfo {author} {\bibfnamefont
  {G.}~\bibnamefont {Niehues}}, \bibinfo {author} {\bibfnamefont
  {M.}~\bibnamefont {Patil}}, \bibinfo {author} {\bibfnamefont
  {B.}~\bibnamefont {Schmitt}}, \bibinfo {author} {\bibfnamefont
  {P.}~\bibnamefont {Sch{\"{o}}nfeldt}}, \bibinfo {author} {\bibfnamefont
  {B.}~\bibnamefont {Steffen}}, \ and\ \bibinfo {author} {\bibfnamefont
  {M.}~\bibnamefont {Weber}},\ }\href {\doibase 10.1016/j.nima.2018.10.093}
  {\bibfield  {journal} {\bibinfo  {journal} {Nuclear Instruments and Methods
  in Physics Research Section A: Accelerators, Spectrometers, Detectors and
  Associated Equipment}\ }\textbf {\bibinfo {volume} {936}},\ \bibinfo {pages}
  {10} (\bibinfo {year} {2019})}\BibitemShut {NoStop}%
\bibitem [{\citenamefont {Casalbuoni}\ \emph {et~al.}(2008)\citenamefont
  {Casalbuoni}, \citenamefont {Schlarb}, \citenamefont {Schmidt}, \citenamefont
  {Schm{\"{u}}ser}, \citenamefont {Steffen},\ and\ \citenamefont
  {Winter}}]{Casalbuoni2008}%
  \BibitemOpen
  \bibfield  {author} {\bibinfo {author} {\bibfnamefont {S.}~\bibnamefont
  {Casalbuoni}}, \bibinfo {author} {\bibfnamefont {H.}~\bibnamefont {Schlarb}},
  \bibinfo {author} {\bibfnamefont {B.}~\bibnamefont {Schmidt}}, \bibinfo
  {author} {\bibfnamefont {P.}~\bibnamefont {Schm{\"{u}}ser}}, \bibinfo
  {author} {\bibfnamefont {B.}~\bibnamefont {Steffen}}, \ and\ \bibinfo
  {author} {\bibfnamefont {A.}~\bibnamefont {Winter}},\ }\href {\doibase
  10.1103/PhysRevSTAB.11.072802} {\bibfield  {journal} {\bibinfo  {journal}
  {Physical Review Special Topics - Accelerators and Beams}\ }\textbf {\bibinfo
  {volume} {11}},\ \bibinfo {pages} {072802} (\bibinfo {year}
  {2008})}\BibitemShut {NoStop}%
\bibitem [{\citenamefont {M{\"{u}}ller}\ \emph {et~al.}(2012)\citenamefont
  {M{\"{u}}ller}, \citenamefont {Peier}, \citenamefont {Schlott}, \citenamefont
  {Steffen}, \citenamefont {Feurer},\ and\ \citenamefont {Kuske}}]{Muller2012}%
  \BibitemOpen
  \bibfield  {author} {\bibinfo {author} {\bibfnamefont {F.}~\bibnamefont
  {M{\"{u}}ller}}, \bibinfo {author} {\bibfnamefont {P.}~\bibnamefont {Peier}},
  \bibinfo {author} {\bibfnamefont {V.}~\bibnamefont {Schlott}}, \bibinfo
  {author} {\bibfnamefont {B.}~\bibnamefont {Steffen}}, \bibinfo {author}
  {\bibfnamefont {T.}~\bibnamefont {Feurer}}, \ and\ \bibinfo {author}
  {\bibfnamefont {P.}~\bibnamefont {Kuske}},\ }\href {\doibase
  10.1103/PhysRevSTAB.15.070701} {\bibfield  {journal} {\bibinfo  {journal}
  {Physical Review Special Topics - Accelerators and Beams}\ }\textbf {\bibinfo
  {volume} {15}},\ \bibinfo {pages} {070701} (\bibinfo {year}
  {2012})}\BibitemShut {NoStop}%
\bibitem [{\citenamefont {Funkner}\ \emph {et~al.}(2019)\citenamefont
  {Funkner}, \citenamefont {Blomley}, \citenamefont {Br{\"{u}}ndermann},
  \citenamefont {Caselle}, \citenamefont {Hiller}, \citenamefont {Nasse},
  \citenamefont {Niehues}, \citenamefont {Rota}, \citenamefont
  {Sch{\"{o}}nfeldt}, \citenamefont {Walther}, \citenamefont {Weber},\ and\
  \citenamefont {M{\"{u}}ller}}]{Funkner2019}%
  \BibitemOpen
  \bibfield  {author} {\bibinfo {author} {\bibfnamefont {S.}~\bibnamefont
  {Funkner}}, \bibinfo {author} {\bibfnamefont {E.}~\bibnamefont {Blomley}},
  \bibinfo {author} {\bibfnamefont {E.}~\bibnamefont {Br{\"{u}}ndermann}},
  \bibinfo {author} {\bibfnamefont {M.}~\bibnamefont {Caselle}}, \bibinfo
  {author} {\bibfnamefont {N.}~\bibnamefont {Hiller}}, \bibinfo {author}
  {\bibfnamefont {M.~J.}\ \bibnamefont {Nasse}}, \bibinfo {author}
  {\bibfnamefont {G.}~\bibnamefont {Niehues}}, \bibinfo {author} {\bibfnamefont
  {L.}~\bibnamefont {Rota}}, \bibinfo {author} {\bibfnamefont {P.}~\bibnamefont
  {Sch{\"{o}}nfeldt}}, \bibinfo {author} {\bibfnamefont {S.}~\bibnamefont
  {Walther}}, \bibinfo {author} {\bibfnamefont {M.}~\bibnamefont {Weber}}, \
  and\ \bibinfo {author} {\bibfnamefont {A.-S.}\ \bibnamefont {M{\"{u}}ller}},\
  }\href {\doibase 10.1103/PhysRevAccelBeams.22.022801} {\bibfield  {journal}
  {\bibinfo  {journal} {Physical Review Accelerators and Beams}\ }\textbf
  {\bibinfo {volume} {22}},\ \bibinfo {pages} {022801} (\bibinfo {year}
  {2019})}\BibitemShut {NoStop}%
\bibitem [{\citenamefont {Buzug}(2008)}]{Buzug2008}%
  \BibitemOpen
  \bibfield  {author} {\bibinfo {author} {\bibfnamefont {T.~M.}\ \bibnamefont
  {Buzug}},\ }\href {\doibase 10.1007/978-3-540-39408-2} {\emph {\bibinfo
  {title} {{Computed Tomography}}}}\ (\bibinfo  {publisher} {Springer Berlin
  Heidelberg},\ \bibinfo {address} {Berlin, Heidelberg},\ \bibinfo {year}
  {2008})\BibitemShut {NoStop}%
\bibitem [{\citenamefont {Wille}(2001)}]{Wille2001}%
  \BibitemOpen
  \bibfield  {author} {\bibinfo {author} {\bibfnamefont {K.}~\bibnamefont
  {Wille}},\ }\href@noop {} {\emph {\bibinfo {title} {{The Physics of Particle
  Accelerators}}}}\ (\bibinfo  {publisher} {Clarendon Press},\ \bibinfo {year}
  {2001})\BibitemShut {NoStop}%
\bibitem [{\citenamefont {J{\"{a}}hne}(2005)}]{Jahne2005}%
  \BibitemOpen
  \bibfield  {author} {\bibinfo {author} {\bibfnamefont {B.}~\bibnamefont
  {J{\"{a}}hne}},\ }\href {\doibase 10.1007/b138991} {\emph {\bibinfo {title}
  {{Digitale Bildverarbeitung}}}},\ \bibinfo {edition} {5th}\ ed.\ (\bibinfo
  {publisher} {Springer-Verlag},\ \bibinfo {address} {Berlin/Heidelberg},\
  \bibinfo {year} {2005})\BibitemShut {NoStop}%
\bibitem [{\citenamefont {Ormond}\ and\ \citenamefont {Rogers}(1997)}]{Ormond}%
  \BibitemOpen
  \bibfield  {author} {\bibinfo {author} {\bibfnamefont {K.}~\bibnamefont
  {Ormond}}\ and\ \bibinfo {author} {\bibfnamefont {J.}~\bibnamefont
  {Rogers}},\ }in\ \href {\doibase 10.1109/PAC.1997.751029} {\emph {\bibinfo
  {booktitle} {Proceedings of the 1997 Particle Accelerator Conference}}},\
  Vol.~\bibinfo {volume} {2}\ (\bibinfo  {publisher} {IEEE},\ \bibinfo
  {address} {Vancouver},\ \bibinfo {year} {1997})\ pp.\ \bibinfo {pages}
  {1822--1824}\BibitemShut {NoStop}%
\bibitem [{\citenamefont {Kehrer}\ \emph {et~al.}(2018)\citenamefont {Kehrer},
  \citenamefont {Brosi}, \citenamefont {Steinmann}, \citenamefont {Blomley},
  \citenamefont {Br{\"{u}}ndermann}, \citenamefont {Caselle}, \citenamefont
  {Funkner}, \citenamefont {Hiller}, \citenamefont {Nasse}, \citenamefont
  {Niehues}, \citenamefont {Rota}, \citenamefont {Schedler}, \citenamefont
  {Sch{\"{o}}nfeldt}, \citenamefont {Schuh}, \citenamefont {Sch{\"{u}}tze},
  \citenamefont {Weber},\ and\ \citenamefont {M{\"{u}}ller}}]{Kehrer2018}%
  \BibitemOpen
  \bibfield  {author} {\bibinfo {author} {\bibfnamefont {B.}~\bibnamefont
  {Kehrer}}, \bibinfo {author} {\bibfnamefont {M.}~\bibnamefont {Brosi}},
  \bibinfo {author} {\bibfnamefont {J.~L.}\ \bibnamefont {Steinmann}}, \bibinfo
  {author} {\bibfnamefont {E.}~\bibnamefont {Blomley}}, \bibinfo {author}
  {\bibfnamefont {E.}~\bibnamefont {Br{\"{u}}ndermann}}, \bibinfo {author}
  {\bibfnamefont {M.}~\bibnamefont {Caselle}}, \bibinfo {author} {\bibfnamefont
  {S.}~\bibnamefont {Funkner}}, \bibinfo {author} {\bibfnamefont
  {N.}~\bibnamefont {Hiller}}, \bibinfo {author} {\bibfnamefont {M.~J.}\
  \bibnamefont {Nasse}}, \bibinfo {author} {\bibfnamefont {G.}~\bibnamefont
  {Niehues}}, \bibinfo {author} {\bibfnamefont {L.}~\bibnamefont {Rota}},
  \bibinfo {author} {\bibfnamefont {M.}~\bibnamefont {Schedler}}, \bibinfo
  {author} {\bibfnamefont {P.}~\bibnamefont {Sch{\"{o}}nfeldt}}, \bibinfo
  {author} {\bibfnamefont {M.}~\bibnamefont {Schuh}}, \bibinfo {author}
  {\bibfnamefont {P.}~\bibnamefont {Sch{\"{u}}tze}}, \bibinfo {author}
  {\bibfnamefont {M.}~\bibnamefont {Weber}}, \ and\ \bibinfo {author}
  {\bibfnamefont {A.-S.}\ \bibnamefont {M{\"{u}}ller}},\ }\href {\doibase
  10.1103/PhysRevAccelBeams.21.102803} {\bibfield  {journal} {\bibinfo
  {journal} {Physical Review Accelerators and Beams}\ }\textbf {\bibinfo
  {volume} {21}},\ \bibinfo {pages} {102803} (\bibinfo {year}
  {2018})}\BibitemShut {NoStop}%
\bibitem [{\citenamefont {Nakhmani}\ and\ \citenamefont
  {Tannenbaum}(2013)}]{Nakhmani2013}%
  \BibitemOpen
  \bibfield  {author} {\bibinfo {author} {\bibfnamefont {A.}~\bibnamefont
  {Nakhmani}}\ and\ \bibinfo {author} {\bibfnamefont {A.}~\bibnamefont
  {Tannenbaum}},\ }\href {\doibase 10.1016/j.patrec.2012.10.025} {\bibfield
  {journal} {\bibinfo  {journal} {Pattern Recognition Letters}\ }\textbf
  {\bibinfo {volume} {34}},\ \bibinfo {pages} {315} (\bibinfo {year}
  {2013})}\BibitemShut {NoStop}%
\bibitem [{\citenamefont {{Liwei Wang}}\ \emph {et~al.}(2005)\citenamefont
  {{Liwei Wang}}, \citenamefont {{Yan Zhang}},\ and\ \citenamefont {{Jufu
  Feng}}}]{LiweiWang2005}%
  \BibitemOpen
  \bibfield  {author} {\bibinfo {author} {\bibnamefont {{Liwei Wang}}},
  \bibinfo {author} {\bibnamefont {{Yan Zhang}}}, \ and\ \bibinfo {author}
  {\bibnamefont {{Jufu Feng}}},\ }\href {\doibase 10.1109/TPAMI.2005.165}
  {\bibfield  {journal} {\bibinfo  {journal} {IEEE Transactions on Pattern
  Analysis and Machine Intelligence}\ }\textbf {\bibinfo {volume} {27}},\
  \bibinfo {pages} {1334} (\bibinfo {year} {2005})}\BibitemShut {NoStop}%
\bibitem [{\citenamefont {Roussel}\ \emph {et~al.}(2014)\citenamefont
  {Roussel}, \citenamefont {Evain}, \citenamefont {Szwaj},\ and\ \citenamefont
  {Bielawski}}]{Roussel2014}%
  \BibitemOpen
  \bibfield  {author} {\bibinfo {author} {\bibfnamefont {E.}~\bibnamefont
  {Roussel}}, \bibinfo {author} {\bibfnamefont {C.}~\bibnamefont {Evain}},
  \bibinfo {author} {\bibfnamefont {C.}~\bibnamefont {Szwaj}}, \ and\ \bibinfo
  {author} {\bibfnamefont {S.}~\bibnamefont {Bielawski}},\ }\href {\doibase
  10.1103/PhysRevSTAB.17.010701} {\bibfield  {journal} {\bibinfo  {journal}
  {Physical Review Special Topics - Accelerators and Beams}\ }\textbf {\bibinfo
  {volume} {17}},\ \bibinfo {pages} {010701} (\bibinfo {year}
  {2014})}\BibitemShut {NoStop}%
\bibitem [{\citenamefont {Seletskiy}\ \emph {et~al.}(2013)\citenamefont
  {Seletskiy}, \citenamefont {Podobedov}, \citenamefont {Shen},\ and\
  \citenamefont {Yang}}]{Seletskiy2013}%
  \BibitemOpen
  \bibfield  {author} {\bibinfo {author} {\bibfnamefont {S.}~\bibnamefont
  {Seletskiy}}, \bibinfo {author} {\bibfnamefont {B.}~\bibnamefont
  {Podobedov}}, \bibinfo {author} {\bibfnamefont {Y.}~\bibnamefont {Shen}}, \
  and\ \bibinfo {author} {\bibfnamefont {X.}~\bibnamefont {Yang}},\ }\href
  {\doibase 10.1103/PhysRevLett.111.034803} {\bibfield  {journal} {\bibinfo
  {journal} {Physical Review Letters}\ }\textbf {\bibinfo {volume} {111}},\
  \bibinfo {pages} {034803} (\bibinfo {year} {2013})}\BibitemShut {NoStop}%
\bibitem [{\citenamefont {M{\"{u}}ller}\ \emph {et~al.}(2005)\citenamefont
  {M{\"{u}}ller}, \citenamefont {Birkel}, \citenamefont {Gasharova},
  \citenamefont {Huttel}, \citenamefont {Kubat}, \citenamefont {Mathis},
  \citenamefont {Moss}, \citenamefont {Mexner}, \citenamefont {Rossmanith},
  \citenamefont {Wuensch}, \citenamefont {Wesolowski}, \citenamefont {Perez},
  \citenamefont {Pont},\ and\ \citenamefont {Hirschmugl}}]{Muller2005}%
  \BibitemOpen
  \bibfield  {author} {\bibinfo {author} {\bibfnamefont {A.-S.}\ \bibnamefont
  {M{\"{u}}ller}}, \bibinfo {author} {\bibfnamefont {I.}~\bibnamefont
  {Birkel}}, \bibinfo {author} {\bibfnamefont {B.}~\bibnamefont {Gasharova}},
  \bibinfo {author} {\bibfnamefont {E.}~\bibnamefont {Huttel}}, \bibinfo
  {author} {\bibfnamefont {R.}~\bibnamefont {Kubat}}, \bibinfo {author}
  {\bibfnamefont {Y.-L.}\ \bibnamefont {Mathis}}, \bibinfo {author}
  {\bibfnamefont {D.}~\bibnamefont {Moss}}, \bibinfo {author} {\bibfnamefont
  {W.}~\bibnamefont {Mexner}}, \bibinfo {author} {\bibfnamefont
  {R.}~\bibnamefont {Rossmanith}}, \bibinfo {author} {\bibfnamefont
  {M.}~\bibnamefont {Wuensch}}, \bibinfo {author} {\bibfnamefont
  {P.}~\bibnamefont {Wesolowski}}, \bibinfo {author} {\bibfnamefont
  {F.}~\bibnamefont {Perez}}, \bibinfo {author} {\bibfnamefont
  {M.}~\bibnamefont {Pont}}, \ and\ \bibinfo {author} {\bibfnamefont
  {C.}~\bibnamefont {Hirschmugl}},\ }in\ \href {\doibase
  10.1109/PAC.2005.1591164} {\emph {\bibinfo {booktitle} {Proceedings of the
  2005 Particle Accelerator Conference}}}\ (\bibinfo  {publisher} {IEEE},\
  \bibinfo {address} {Knoxville},\ \bibinfo {year} {2005})\ pp.\ \bibinfo
  {pages} {2518--2520}\BibitemShut {NoStop}%
\bibitem [{\citenamefont {Sch{\"{o}}nfeldt}\ \emph
  {et~al.}(2017{\natexlab{b}})\citenamefont {Sch{\"{o}}nfeldt}, \citenamefont
  {Blomley}, \citenamefont {Br{\"{u}}ndermann}, \citenamefont {Caselle},
  \citenamefont {Funkner}, \citenamefont {Hiller}, \citenamefont {Kehrer},
  \citenamefont {Nasse}, \citenamefont {Niehues}, \citenamefont {Rota},
  \citenamefont {Schedler}, \citenamefont {Schuh}, \citenamefont {Weber},\ and\
  \citenamefont
  {M{\"{u}}ller}}]{SchoenfeldtBlomleyBruendermann2017_1000071492}%
  \BibitemOpen
  \bibfield  {author} {\bibinfo {author} {\bibfnamefont {P.}~\bibnamefont
  {Sch{\"{o}}nfeldt}}, \bibinfo {author} {\bibfnamefont {E.}~\bibnamefont
  {Blomley}}, \bibinfo {author} {\bibfnamefont {E.}~\bibnamefont
  {Br{\"{u}}ndermann}}, \bibinfo {author} {\bibfnamefont {M.}~\bibnamefont
  {Caselle}}, \bibinfo {author} {\bibfnamefont {S.}~\bibnamefont {Funkner}},
  \bibinfo {author} {\bibfnamefont {N.}~\bibnamefont {Hiller}}, \bibinfo
  {author} {\bibfnamefont {B.}~\bibnamefont {Kehrer}}, \bibinfo {author}
  {\bibfnamefont {M.~J.}\ \bibnamefont {Nasse}}, \bibinfo {author}
  {\bibfnamefont {G.}~\bibnamefont {Niehues}}, \bibinfo {author} {\bibfnamefont
  {L.}~\bibnamefont {Rota}}, \bibinfo {author} {\bibfnamefont {M.}~\bibnamefont
  {Schedler}}, \bibinfo {author} {\bibfnamefont {M.}~\bibnamefont {Schuh}},
  \bibinfo {author} {\bibfnamefont {M.}~\bibnamefont {Weber}}, \ and\ \bibinfo
  {author} {\bibfnamefont {A.~S.}\ \bibnamefont {M{\"{u}}ller}},\ }in\ \href
  {\doibase 10.18429/JACoW-IPAC2017-MOPAB055} {\emph {\bibinfo {booktitle}
  {Proceedings of the 8th International Particle Accelerator Conference,
  Copenhagen, Denmark}}}\ (\bibinfo  {publisher} {JACoW},\ \bibinfo {year}
  {2017})\ pp.\ \bibinfo {pages} {227--230}\BibitemShut {NoStop}%
\bibitem [{\citenamefont {Haissinski}(1973)}]{haissinski1973}%
  \BibitemOpen
  \bibfield  {author} {\bibinfo {author} {\bibfnamefont {J.}~\bibnamefont
  {Haissinski}},\ }\href@noop {} {\bibfield  {journal} {\bibinfo  {journal} {Il
  Nuovo Cimento B}\ }\textbf {\bibinfo {volume} {18}},\ \bibinfo {pages} {72}
  (\bibinfo {year} {1973})}\BibitemShut {NoStop}%
\bibitem [{\citenamefont {Ng}(2006)}]{ng_2006}%
  \BibitemOpen
  \bibfield  {author} {\bibinfo {author} {\bibfnamefont {K.~Y.}\ \bibnamefont
  {Ng}},\ }\href {\doibase 10.1142/5835} {\emph {\bibinfo {title} {{Physics of
  Intensity Dependent Beam Instabilities}}}}\ (\bibinfo  {publisher} {World
  Scientific Publishing},\ \bibinfo {year} {2006})\BibitemShut {NoStop}%
\bibitem [{\citenamefont {Krinsky}\ and\ \citenamefont
  {Wang}(1985)}]{Krinsky:1985sq}%
  \BibitemOpen
  \bibfield  {author} {\bibinfo {author} {\bibfnamefont {S.}~\bibnamefont
  {Krinsky}}\ and\ \bibinfo {author} {\bibfnamefont {J.~M.}\ \bibnamefont
  {Wang}},\ }\href@noop {} {\bibfield  {journal} {\bibinfo  {journal} {Part.
  Accel.}\ }\textbf {\bibinfo {volume} {17}},\ \bibinfo {pages} {109} (\bibinfo
  {year} {1985})}\BibitemShut {NoStop}%
\bibitem [{\citenamefont {Sch{\"{o}}nfeldt}\ \emph {et~al.}(2018)\citenamefont
  {Sch{\"{o}}nfeldt}, \citenamefont {Boltz}, \citenamefont {Mochihashi},
  \citenamefont {Steinmann},\ and\ \citenamefont
  {M{\"{u}}ller}}]{Schonfeldt2018a}%
  \BibitemOpen
  \bibfield  {author} {\bibinfo {author} {\bibfnamefont {P.}~\bibnamefont
  {Sch{\"{o}}nfeldt}}, \bibinfo {author} {\bibfnamefont {T.}~\bibnamefont
  {Boltz}}, \bibinfo {author} {\bibfnamefont {A.}~\bibnamefont {Mochihashi}},
  \bibinfo {author} {\bibfnamefont {J.~L.}\ \bibnamefont {Steinmann}}, \ and\
  \bibinfo {author} {\bibfnamefont {A.-S.}\ \bibnamefont {M{\"{u}}ller}},\
  }\href {\doibase 10.1088/1742-6596/1067/6/062025} {\bibfield  {journal}
  {\bibinfo  {journal} {Journal of Physics: Conference Series}\ }\textbf
  {\bibinfo {volume} {1067}},\ \bibinfo {pages} {062025} (\bibinfo {year}
  {2018})}\BibitemShut {NoStop}%
\bibitem [{\citenamefont {Murphy}\ \emph {et~al.}(1997)\citenamefont {Murphy},
  \citenamefont {Krinsky},\ and\ \citenamefont {Gluckstern}}]{murphy1997}%
  \BibitemOpen
  \bibfield  {author} {\bibinfo {author} {\bibfnamefont {J.~B.}\ \bibnamefont
  {Murphy}}, \bibinfo {author} {\bibfnamefont {S.}~\bibnamefont {Krinsky}}, \
  and\ \bibinfo {author} {\bibfnamefont {R.~L.}\ \bibnamefont {Gluckstern}},\
  }\href {http://cds.cern.ch/record/1120287/files/p9.pdf} {\bibfield  {journal}
  {\bibinfo  {journal} {Particle Accelerators}\ }\textbf {\bibinfo {volume}
  {57}},\ \bibinfo {pages} {9} (\bibinfo {year} {1997})}\BibitemShut {NoStop}%
\end{thebibliography}%
 
\end{document}